\documentclass[a4paper]{article}

\usepackage[margin=1in]{geometry} 

\usepackage{amsmath}
\usepackage{amsthm}
\usepackage{amssymb}

\usepackage[utf8]{inputenc}
\usepackage[hidelinks]{hyperref}
\hypersetup{
	unicode,
	pdfauthor={Author One, Author Two, Author Three},
	pdftitle={A simple article template},
	pdfsubject={A simple article template},
	pdfkeywords={article, template, simple},
	pdfproducer={LaTeX},
	pdfcreator={pdflatex}
}

\usepackage[sort&compress,numbers,square]{natbib}
\theoremstyle{plain}

\theoremstyle{definition}

\usepackage{graphicx, color}
\graphicspath{{fig/}}

\usepackage{comment}
\usepackage{color}

\usepackage{algorithm, algpseudocode} 
\usepackage{mathrsfs} 

\usepackage{lipsum}

\title{Inverse scattering transform for continuous and discrete space-time shifted integrable equations}
\author{Mark J. Ablowitz$^1$ \and Ziad H. Musslimani$^2$ \and Nicholas J. Ossi$^{2\dagger}$}

\date{
	$^1$ Department of Applied Mathematics, University of Colorado, Campus Box 526, Boulder, Colorado 80309-0526\\
	$^2$ Department of Mathematics, Florida State University, Tallahassee, FL 32306-4510\\
 $\dagger$ Corresponding author: nossi@math.fsu.edu}

\begin{document}
	\maketitle
	\begin{abstract}
Nonlocal integrable partial differential equations possessing a spatial or temporal reflection have constituted an active research area for the past decade. Recently, more general classes of these nonlocal equations have been proposed, wherein the nonlocality appears as a combination of a shift (by a real or a complex parameter) and a reflection. This new shifting parameter manifests itself in the inverse scattering transform (IST) as an additional phase factor in an analogous way to the classical Fourier transform. In this paper, the IST is analyzed in detail for several examples of such systems. Particularly, time, space, and space-time shifted nonlinear Schr\"odinger (NLS) and space-time shifted modified Korteweg-de Vries (mKdV) equations are studied. Additionally, the semi-discrete IST is developed for the time, space and space-time shifted variants of the Ablowitz-Ladik integrable discretization of the NLS. One-soliton solutions are constructed for all continuous and discrete cases.
	\end{abstract}
\tableofcontents
	\section{Introduction}
	\subsection{Backround}
\label{intro}
The study of infinite-dimensional integrable evolution equations began in 1967 
when a method (based on inverse scattering) to exactly solve the initial value problem associated with the Korteweg-de Vries (KdV) equation on the line was presented \cite{GGKM}. Soon thereafter, Lax presented an alternative approach to integrate the KdV equation by reformulating it as a compatibility condition of two linear equations (one of which is of a Schr\"odinger equation type) \cite{lax}. Inspired by Lax's ideas, the nonlinear Schr\"odinger (NLS) equation, which finds applications in fluid mechanics, Bose-Einstein condensation, and optics (to mention a few) \cite{Fibich}, was also exactly solved \cite{ZS}.
In 1974, Ablowitz, Kaup, Newell, and Segur put forward more general classes of integrable nonlinear partial differential equations solvable by the so-called inverse scattering transform (IST) \cite{AKNS}. In their work, the IST was applied to extensive hierarchies of integrable nonlinear evolution equations, which include additional physically significant models such as the modified KdV (mKdV) and sine-Gordon (SG) equations. For comprehensive discussions
of the IST and its applications in the theory of nonlinear waves, see \cite{APT_Book, Ablowitz2, Ablowitz1, Abl2011}.

Several decades later, a new \textit{nonlocal} integrable equation was introduced in 2013 \cite{pt2013}. It is a spatially nonlocal variant of the NLS equation that possesses PT-symmetry (invariance under the combined action of spatial inversion and complex conjugation, see \cite{ptsym}). The discovery of this system
was just the beginning of an influx of novel integrable partial differential equations, including nonlocal variants of the mKdV, SG, and many more. The integrable structure of these nonlocal models has been exploited to obtain exact solutions with zero \cite{AM_Nonlinearity,AM_Studies}, nonzero \cite{AblowitzLuoMusslimani1,AblowitzLuoMusslimani4}, and periodic \cite{periodic} boundary conditions. Furthermore, the Ablowitz-Ladik (AL) semi-discrete system, an integrable spatially discretized version of the NLS \cite{AL1,AL2,Kevrekidis}, was shown to have nonlocal reductions as well, for which the IST is explored in \cite{AM_discrete}. 

In 2021 \cite{AM_shifted}, several new space-time {\it shifted} nonlocal integrable symmetry reductions to 
the well-known AKNS $(q,r)$ system \cite{AKNS}, 
\begin{eqnarray}
\label{q-sys}
iq_t(x,t) &=& q_{xx}(x,t) - 2q^2(x,t)r(x,t),\\
\label{r-sys}
-ir_t(x,t) &=& r_{xx}(x,t) - 2r^2(x,t)q(x,t),
\end{eqnarray}
were identified. They are of the form
\begin{eqnarray}
\label{shift-x-reduction}
 r(x,t) &=& \sigma q^*(x_0-x,t),\;x_{0}\in\mathbb{R},\\
\label{shift-t-only-reduction}
 r(x,t) &=& \sigma q(x,t_0-t),\;t_{0}\in\mathbb{C},\\
\label{shift-x-t-reduction}
 r(x,t) &=& \sigma q(x_0-x,t_0-t),\;x_{0},t_{0}\in\mathbb{C},
\end{eqnarray}
where $\sigma = \pm 1$ and $x_0,t_0$ are arbitrary parameters. Under these integrable symmetry reductions, (\ref{q-sys})-(\ref{r-sys}) are compatible and 
lead to the so-called space, time, and space-time shifted integrable nonlocal NLS equations 
respectively, given by:
\begin{eqnarray}
\label{shifted-PT-nonloc-NLS}
iq_t(x,t) &=& q_{xx}(x,t) - 2\sigma q^2(x,t)q^*(x_0-x,t),\\
\label{shifted-t-only-nonloc-NLS}
iq_t(x,t) &=& q_{xx}(x,t) - 2\sigma q^2(x,t)q(x,t_0-t),\\
\label{shifted-x-t-nonloc-NLS}
iq_t(x,t) &=& q_{xx}(x,t) - 2\sigma q^2(x,t)q(x_0-x,t_0-t).
\end{eqnarray}
The inverse scattering transform for the space-shifted NLS equation, i.e. Eq. \eqref{shifted-PT-nonloc-NLS}, was briefly outlined in \cite{AM_shifted}. The one and two soliton solutions to \eqref{shifted-PT-nonloc-NLS} were obtained, and its relation to the unshifted ($x_{0}=0$) case was also discussed. Exact solutions for many of these shifted equations have also been obtained through various direct methods (see for example \cite{shift1,shift2,shift3,shift4,shift5,shift6,shift7,shift8}).

Furthermore, in \cite{AM_shifted} it was shown that the $(q,r)$ system associated with the mKdV equation, i.e.
\begin{eqnarray}
\label{q-sys-mkdv}
q_t(x,t) + q_{xxx}(x,t) - 6q(x,t)r(x,t)q_x(x,t) &=& 0,\\
\label{r-sys-mkdv}
r_t(x,t) + r_{xxx}(x,t) - 6q(x,t)r(x,t)r_x(x,t) &=& 0,
\end{eqnarray}
also admits similar shifted symmetry reductions. Specifically,
\begin{eqnarray}
\label{shift-x-t-reduction-mdkv}
 r(x,t) &=& \sigma q(x_0-x,t_0-t),\;x_{0},t_{0}\in\mathbb{C},\\
 \label{shift-x-t-complex-reduction}
 r(x,t) &=& \sigma q^*(x_0-x,t_0-t),\;x_{0},t_{0}\in\mathbb{R},
\end{eqnarray}
reduce the coupled evolution equations (\ref{q-sys-mkdv}) and  (\ref{r-sys-mkdv}) to the space-time shifted mKdV equation and complex conjugate space-time shifted mKdV equation respectively:
\begin{eqnarray}
\label{space-time-shifted-real-mkdv}
&&q_t(x,t) + q_{xxx}(x,t) - 6\sigma q(x,t) q(x_0-x,t_0-t) q_x(x,t) = 0,\\
\label{space-time-shifted-complex-mkdv}
&&q_t(x,t) + q_{xxx}(x,t) - 6\sigma q(x,t) q^*(x_0-x,t_0-t) q_x(x,t) = 0.
\end{eqnarray}
\subsection{Applications, connection to PT physics, and complex reductions}
In this section, we briefly mention some motivation and potential areas of application for the integrable shifted models considered in this paper.    
\begin{itemize}
\item {\it Nonlinear Klein-Gordon equation}. \\\\
The well-known AKNS $(q,r)$ system (\ref{q-sys})-(\ref{r-sys}) forms a cornerstone system in integrability theory \cite{AKNS}. 
In this regard, an important question is whether one can derive these equations
from a physical system. In \cite{AbMu2019} it was shown the the AKNS $(q,r)$ system can be obtained from the water wave equations, the KdV equation, and the nonlinear Klein-Gordon equation (KG); we will  use  the KG equation for ease of presentation.
Consider the nonlinear KG equation
\begin{eqnarray}
\label{KG-eqn}
\phi_{tt} - \phi_{xx}  + \phi - \phi^3 = 0.
\end{eqnarray}
Introduce new fast and slow scales: $\theta = kx-\omega t,\; X=\epsilon x,\; T=\epsilon t$ with the small 
parameter $|\epsilon| \ll1$ and let $\phi = \phi (\theta, X,T;\epsilon).$
Now assume a perturbative solution to (\ref{KG-eqn}) in the form
\begin{eqnarray}
\label{KG-sol}
\phi = \epsilon \phi_0 +  \epsilon^2 \phi_1 +\cdots .
\end{eqnarray}
The leading order equation reads
\begin{eqnarray}
\label{KG-sol2}
(\partial_\theta^2 + 1)  \phi_0 =0,
\end{eqnarray}
whose solution is given by
\begin{eqnarray}
\label{KG-sol3}
 \phi_0 = A(X,T)e^{i\theta} + B(X,T)e^{-i\theta} ,
\end{eqnarray}
Note that we do not require $B=A^*.$ Removing secular terms that arise at the next order in $\epsilon$ leads to the system of equations given in (\ref{q-sys}) and (\ref{r-sys}) with $q,r$ being proportional to $A$ and $B$ respectively. From the $q,r$ system we can derive the shifted equations  (\ref{shifted-PT-nonloc-NLS}-\ref{shifted-x-t-nonloc-NLS}) mentioned above.
We remark that: i) After rescaling, these secular equations yield the $q,r$ system; ii) The $q,r$ system can be derived in a similar way from the KdV and classical water wave equations.
\item {\it Six wave interaction equations in finite depth gravity waves with surface tension.}\\\\
In \cite{AbFoMu2006}  it was shown that the classical water wave equations with surface tension can be written in terms of the following convenient nonlocal  system
\begin{equation}
 \int_{\mathbb{R}^2} d {\bf r} e^{ - i {\bf k} \cdot {\bf r} }
  \left( i \eta_t \cosh[k (\eta+h)] 
  -  \frac{{\bf k} \cdot \nabla q}{k}  \sinh \left[ k (\eta+h)\right]\right) = 0 ,
  \label{nlcl1}
\end{equation}
 \begin{equation}
 q_t + \frac{1}{2} | \nabla q|^2 + g\eta - \frac{(\eta_t + \nabla q \cdot \nabla \eta)^2} {2(1+|\nabla \eta|^2)}=
\sigma \nabla \cdot \bigg( \frac{\nabla \eta}{\sqrt{1+|\nabla \eta|^2}}\bigg)  ,
\label{nlcl2}
\end{equation}
where $\eta ({\bf r} ,t)$ is the surface elevation and $q=q({\bf r},t)= \phi({\bf r},\eta({\bf r},t),t)$ where  $\phi$ is the velocity potential.
In \cite{AbLuMu2023} it was shown that the leading order solution to system (\ref{nlcl1}) and (\ref{nlcl2}) in a multiscale perturbation expansion 
is given by
\begin{equation}
\label{eta-three-waves}
\eta^{(0)}({\bf r} , t; {\bf R} , T) = \sum_{j=1}^3 \left( A_j ({\bf R},T) e^{-i \theta_j} +B_j({\bf R},T) e^{i \theta_j}  \right),
\end{equation}
\begin{equation}
\label{Q-three-waves}
{\bf Q}^{(0)}({\bf r} ,t,{\bf R},T)  
=  \sum_{j=1}^3 ~ \frac{ \omega_j {\bf k}_j }{k_j} ~ \left (    A_j({\bf R},T)  e^{-i \theta_j}  
+  B_j({\bf R},T)   e^{i \theta_j}  \right) .
\end{equation}
where ${\bf Q}^{(0)} \equiv \nabla q^{(0)}$, $\theta_j \equiv {\bf k}_j \cdot {\bf r}  - \omega_j t, ~~\omega_j=\omega(|{\bf k}_j|), {\bf R}=\epsilon {\bf r}, T=\epsilon t.$ Again, we do not require $B_j({\bf R},T) = A^*_j({\bf R},T).$
To remove resonant terms at the next order in $\epsilon$ we find that the amplitudes $A_j, B_j$ satisfy the following six wave  interaction equations:
\begin{align}
\label{3WEqFa}
&  (\partial_T+C_{1,x}\partial_x + C_{1,y}\partial_y)A_1 - i \sigma_1 B_2A_3 = 0  , \\
\nonumber
&  (\partial_T+C_{2,x}\partial_x + C_{2,y}\partial_y)A_2 - i \sigma_2 B_1A_3 = 0 , \\
\nonumber
& (\partial_T+C_{3,x}\partial_x+ C_{3,y}\partial_y)A_3 -  i \sigma_3 A_1A_2 = 0 , \\
\nonumber
&  (\partial_T+C_{1,x}\partial_x+ C_{1,y}\partial_y)B_1 + i \sigma_1 A_2B_3 = 0 , \\
\nonumber
& (\partial_T+C_{2,x}\partial_x+ C_{2,y}\partial_y)B_2 +  i \sigma_2 A_1B_3 = 0 , \\
\nonumber
& (\partial_T+C_{3,x}\partial_x+ C_{3,y}\partial_y)B_3  +  i\sigma_3 B_1B_2 = 0 , 
\end{align}
where $C_{j,x},C_{j,y}, \sigma_j, j=1,2,3$ are constants was derived in the context of $3\times3$  linear eigenvalue problem
by Ablowitz and Haberman \cite{AbHa1975}. This is the higher order analog of the $q,r$ system (\ref{q-sys})-(\ref{r-sys}) derived by AKNS from a $2 \times 2$  eigenvalue problem. The above six-wave system has reductions to the following 
integrable three-wave equations:
\begin{itemize}
\item classical local:  $B_j(x,t) = A_j^{*}(x,t),\; j=1,2,3.$ 
\item shifted complex nonlocal: $B_j(x,t)=A_j^{*}(x_0-x,t_0-t),\; j=1,2,3,\; x_0,t_0\in\mathbb{R}.$
\item shifted real nonlocal: $B_j(x,t)=A_j(x_0-x,t_0-t),\; j=1,2,3,\; x_0,t_0\in\mathbb{C}$.
\\
\end{itemize}
\item {\it Wave propagation in shifted two-fold PT-symmetric optical lattices.}\\\\
The fundamental connection between all the shifted nonlocal NLS equations, e.g. (\ref{shifted-PT-nonloc-NLS}), and wave propagation in photonic PT-symmetric systems originates from the introduction of the nonlinearly induced potential
\begin{eqnarray}
\label{induced-pot}
V(x,t)=q(x,t)q^*(x_0-x,t).
\end{eqnarray}
Such identification, allows one to view Eq.~(\ref{shifted-PT-nonloc-NLS}) as a ``paraxial wave equation" governing the propagation of an optical beam inside a nonlinearly induced complex refractive index $V$. Note that, in our case, this optical potential satisfies the following shifted PT symmetry
\begin{eqnarray}
\label{induced-pot-sym}
V(x,t)=V^*(x_0-x,t).
\end{eqnarray}
PT-symmetric and non-Hermitian photonic systems have been the subject of intense research for the last decade 
\cite{ZHM1,ZHM2,ZHM3,ZHM4,PT_Yang}. In \cite{CMMCR2016} novel non-Hermitian photonic potentials whose center of PT symmetry is not around zero were introduced. Such refractive indices satisfy shifted PT symmetry like the one given in (\ref{induced-pot-sym}). It was further shown in \cite{CMMCR2016} that the physics stemming from a non-zero reflection point is different than in the case when the potential has zero reflection point. 
\item {\it Complex reductions of physical equations.}\\\\
It is widely understood that complex reductions of physical problems are very important. For example, self-similar reductions of many evolution equations lead to Painlev\'e type equations (see e.g. \cite{Ablowitz2,Ablowitz1}). The underlying basis of these equations are their properties   in the complex plane, i.e. no moveable critical points.  Two other well-known examples are the self-dual Yang Mills equations which are:
i) reductions of  the Yang-Mills equations \cite{Ward1977} and ii) the self-dual Einstein equations which are reductions of  the Einstein equations \cite{WoodMason1988};
see also \cite{Ablowitz1}.
Indeed, a further reduction of the self-dual Einstein equations are the one dimensional self-dual reductions of the Bianchi IX cosmological models \cite{GibPope1979}. These reductions lead to the Darboux-Halphen equations from which the Chazy equation follows \cite{AbChCl1990}. An important feature of the  Darboux-Halphen-Chazy equations is that they exhibit natural barriers where the solution is analytic inside and outside a moveable circle \cite{AbFo2003} but there is no analytic continuation between these regions. This remarkable result gives support for separate cosmological structures that do not communicate with each other.
\end{itemize}
\subsection{Organization}
In this paper, we present a detailed analysis of the inverse scattering transform associated with all the above symmetry reductions associated with the NLS and mKdV systems. This process involves three major steps: (i) the study of a direct scattering problem, which involves identifying certain crucial symmetries; (ii) determining the time evolution of the so-called scattering data; and (iii) solving an inverse scattering problem via a Riemann-Hilbert formulation. Our study of shifted continuous NLS equations is also extended to the discrete domain (continuous in time and discrete in space), where a detailed presentation of the inverse scattering transform associated with the shifted reductions of the Ablowitz-Ladik hierarchy is outlined. One-soliton solutions for the continuous and discrete shifted cases are given and their properties are studied. Additionally, in the study of classical Fourier transforms, an important general result is that a shift in physical space manifests as a phase shift in Fourier space. We find a similar result in the scattering theory associated these shifted nonlinear equations. \\ 
The paper is organized as follows:
\begin{itemize}
    \item In section 2, we discuss the connection between the shifted nonlocal integrable equations of the present work and their unshifted counterparts. The translational and scaling ``pseudo-invariances" of these shifted equations are highlighted.
    \item In section 3, we summarize the direct scattering problem for the general AKNS scattering problem associated with the NLS and mKdV coupled $(q,r)$ systems. The phase shift formulae are discussed  in section 3.3.
    \item In section 4, we provide detailed derivations of the symmetries of the scattering data induced by each of the shifted nonlocal reductions presented above.
    \item In section 5, we summarize the inverse scattering problem for the general AKNS scattering problem, and discuss the time evolution of the scattering data in both the NLS and mKdV cases.
    \item In section 6, we provide detailed derivations of certain explicit representations of the scattering data that are useful in constructing soliton solutions.
    \item In section 7, we construct the one-soliton solution for each of the shifted nonlocal NLS and mKdV variants. 
    \item In sections 8-12, we repeat all of the analyses of sections 3-7 in the context of shifted nonlocal semi-discrete NLS (Ablowitz-Ladik) systems. In section 8.3 the discrete analogue of the phase shift formula derived for the continuous problem is obtained.
    \item In section 13, we offer some concluding remarks and suggestions for future exploration.
\end{itemize}
\section{Connection between shifted and unshifted nonlocal equations}
\subsection{Continuous systems}
\subsubsection{A family of nonlocal equations with space-time pseudo-translation symmetries}
The nonlocal integrable systems with temporal and/or spatial shifts (like those studied in this paper) provide an important and natural generalization of the nonlocal equations with temporal and/or spatial reflection that have been studied in recent years (precisely the ones without any shifts) \cite{AM_Nonlinearity}. They provide two important physical and mathematical aspects that the unshifted system lacks: (i) space and/or time translation symmetries and (ii) a nonlinearly induced complex PT symmetric potentials with a non-zero reflection symmetry point (like the case studied in \cite{CMMCR2016} where a PT symmetric complex optical lattice  
with shifted two-fold symmetry points were introduced and studied -- see Sec. 1.2 for more details). In this section, we focus our discussion on the crucial role that the shift parameters $x_0$ and $t_0$ play in the restoration of a broken space-time translation symmetries.\\\\
The classical (local) nonlinear Schr\"odinger equation
\begin{equation}
\label{classical-NLS}
iq_t(x,t) = q_{xx}(x,t) - 2\sigma q^2(x,t)q^*(x,t),\;\;\; \sigma = \pm 1,
\end{equation}
admits an important translational invariance symmetry given by the following:
 If $q(x,t)$ is a solution to Eq.~(\ref{classical-NLS}) so is $q(x-x_0/2,t-t_0/2)$ for any real constants $x_0,t_0.$ This statement has two consequences: (i) Space translation invariance allows for the construction of a larger class of solutions with additional free parameters. (ii) Time translation invariance symmetry implies that one does {\it not} need to specify the initial data at time zero, i.e.,  $q(x,0)$. Instead, one can supplement Eq.~(\ref{classical-NLS}) with an initial condition at any suitable ``initial time" while preserving its 
well-posedness.\\\\
 In contrast, the PT symmetric, reverse-time only and reverse space-time nonlocal equations  
 (here $\sigma = \pm 1$)
\begin{eqnarray}
\label{unshifted-PT-nonloc-NLS}
i \partial_{t} q_{\text{un}}(x,t)&=& 
\partial_{x}^2 q_{\text{un}}(x,t) - 2\sigma q^2_{\text{un}}(x,t) q^*_{\text{un}}(-x,t) \;,\\
\label{unshifted-t-nonloc-NLS}
i \partial_{t} q_{\text{un}}(x,t)&=& 
\partial_{x}^2 q_{\text{un}}(x,t) - 2\sigma q^2_{\text{un}}(x,t) q_{\text{un}}(x,-t) ,\\
\label{unshifted-xt-nonloc-NLS}
i \partial_{t} q_{\text{un}}(x,t) &=& 
\partial_{x}^2 q_{\text{un}}(x,t) - 2\sigma q^2_{\text{un}}(x,t) q_{\text{un}}(-x,-t) ,
\end{eqnarray}
respectively break space, time, and space-time translation symmetry on the real line. That is to say:
\begin{equation}
\label{translation-sym1}
\text{If~} q_{\text{un}}(x,t)~ \text{solves~} (\ref{unshifted-PT-nonloc-NLS}) \text{~then~}
q_{\text{un}}(x-x_0/2,t) \text{~does~not~satisfy~} (\ref{unshifted-PT-nonloc-NLS}), x_0\in\mathbb{R}. 
\end{equation}
This means, despite Eq.~(\ref{unshifted-PT-nonloc-NLS}) being a constant coefficient evolution equation, it breaks space translation invariance on the real line. Instead, this symmetry is preserved on the imaginary axis. Indeed, we have:
\begin{equation}
\label{translation-sym2}
\text{If~} q_{\text{un}}(x,t)~ \text{is~a~solution~to~} (\ref{unshifted-PT-nonloc-NLS}) \text{~so~is~}
q_{\text{un}}(x-ix_0/2,t),\; x_0\in\mathbb{R}. 
\end{equation}
Since Eq.~(\ref{unshifted-PT-nonloc-NLS}) is an integrable system, it admits infinitely many conservation laws, two of which are the energy (corresponding to time translation invariance) 
and momentum (related to space translation symmetry). Questions that immediately arise are: 
Why does Eq.~(\ref{unshifted-PT-nonloc-NLS}) break translation invariance on the real line and is there a ``hidden" symmetry associated with it that would restore that broken symmetry? The answer is connected with the shift parameter $x_0$. In fact, we have the following statement (which is further explained in the next subsection):
\begin{equation}
\label{translation-sym3}
\text{If~} q_{\text{un}}(x,t)~ \text{solves~} (\ref{unshifted-PT-nonloc-NLS}) \text{~then~}
q_{\text{un}}(x-x_0/2,t)~\text{is~a~solution~to~the~shifted~nonlocal~NLS~} (\ref{shifted-PT-nonloc-NLS}),\; 
x_0\in\mathbb{R}. 
\end{equation}
Thus, the statement given in (\ref{translation-sym3}) implies the following important result:\\\\
{\it The shifted NLS Eq.~(\ref{shifted-PT-nonloc-NLS}) represents a one-parameter family of nonlocal integrable equations with 
endowed pseudo-translation symmetry in the following sense: if $q(x,t)$ is a flow that belongs to the family, so is $q(x-x_0/2,t)$.}
\\\\
The situation is even more intricate when dealing with Eq.~(\ref{unshifted-t-nonloc-NLS}), where now time translation symmetry is broken both on the real and the imaginary axis. This critical fact again raises the question as to what causes the loss of such 
time-invariance? The resolution to this issue is provided by the introduction of the time shift parameter $t_0$ which endows  
\eqref{shifted-t-only-nonloc-NLS} with a pseudo-time translation symmetry.\\
\\It is also crucial to note that the temporal shifting parameter $t_{0}$ restricts the time at which initial data should be provided. Namely, the time-shifted equations \eqref{shifted-t-only-nonloc-NLS}, \eqref{shifted-x-t-nonloc-NLS}, \eqref{space-time-shifted-real-mkdv}, and \eqref{space-time-shifted-complex-mkdv} are only well-posed if an initial condition is given specifically at time $t=t_{0}/2$. This is in contrast to 
time-local integrable systems (including \eqref{shifted-PT-nonloc-NLS}), for which the well-posedness of initial value problems does not depend on the choice of the initial instant.

\subsubsection{Relations between solution sets of shifted and unshifted nonlocal equations}
As we show below, the solution set of a given shifted integrable equation is related to the solution set of its corresponding unshifted counterpart via a simple transformation. For example, suppose that $q_{\text{un}}(x,t)$ is a given solution of the \textit{unshifted} 
PT symmetric nonlocal NLS Eq.~(\ref{unshifted-PT-nonloc-NLS}). Then, one can show that $q(x,t)$ defined by
\begin{equation}
\label{connection}
    q(x,t)\equiv q_{\text{un}}(x-x_0/2,t),
\end{equation}
is a solution of Eq. \eqref{shifted-PT-nonloc-NLS}. Thus, this transformation maps a solution of the unshifted nonlocal NLS to a solution of the shifted nonlocal NLS. 
Similar statements can be made regarding Eqns.~(\ref{shifted-t-only-nonloc-NLS}) and (\ref{shifted-x-t-nonloc-NLS}). That is to say:
\begin{itemize}
\item
Let $q_{\text{un}}(x,t)$ be a solution of the \textit{unshifted} reverse-time only nonlocal NLS Eq.~(\ref{unshifted-t-nonloc-NLS}). 
Then, 
\begin{equation}
\label{connection-time2}
    q(x,t) \equiv q_{\text{un}}(x,t-t_0/2),
\end{equation}
is a solution to the time only shifted NLS Eq. (\ref{shifted-t-only-nonloc-NLS}).
\item
Let $q_{\text{un}}(x,t)$ be a solution of the \textit{unshifted} reverse space-time nonlocal NLS Eq. (\ref{unshifted-xt-nonloc-NLS}). 
Then, 
\begin{equation}
\label{connection-time2a}
    q(x,t) \equiv q_{\text{un}}(x-x_0/2,t-t_0/2),
\end{equation}
is a solution to the space-time shifted NLS Eq. (\ref{shifted-x-t-nonloc-NLS}).
\end{itemize}
At this point we make two important remarks.   
\begin{enumerate}
\item
 The shift relation (\ref{connection}) provides a direct connection between solutions of Eqns. (\ref{unshifted-PT-nonloc-NLS}) 
 and (\ref{shifted-PT-nonloc-NLS}).   As an example, a pure soliton solution 
of Eq. (\ref{unshifted-PT-nonloc-NLS}) 
yields a soliton solution to Eq. (\ref{shifted-PT-nonloc-NLS}) via  the shift relation (\ref{connection}).
On the other hand, as we have found in \cite{AM_shifted}, 
solving the linear system of equations from IST for a pure soliton solution does not necessarily yield the soliton solution obtained from the shift relation (\ref{connection}). The reason is that there are a family of soliton solutions to Eq. (\ref{unshifted-PT-nonloc-NLS}) due to the fact that there is freedom of choice of norming constants which depend on the shift parameter $x_0$. There will be a suitable choice of norming constants that satisfy equation (\ref{connection}) but they may not come out directly from IST. 
For example, the solutions given in Fig. 1 of  \cite{AM_shifted} obtained from IST for different  value of $x_0$ are all pure soliton solutions but they differ in the time where the solitons interact or can be viewed as distributed differently at, e.g., large negative times. This observation has been verified numerically.
\item There is also a direct transformation  between two shifted equations (with different shifts). Namely if $q(x,t)$ satisfies
\begin{equation}
\label{shiftNLS}
iq_t(x,t) = q_{xx}(x,t) - 2\sigma q^2(x,t)q^*(x_0-x,t) ,
\end{equation}
then $\tilde{q}(x,t)=q(x+\frac{1}{2}(x_0-\tilde{x}_0),t)$ satisfies
\begin{equation}
\label{tildeshiftNLS}
i\tilde{q}_t(x,t) = \tilde{q}_{xx}(x,t) - 2\sigma \tilde{q}^2(x,t)\tilde{q}^*(\tilde{x}_0-x,t) .
\end{equation}
This property highlights the nontrivial role that the shift parameters play in restoring an otherwise broken space translation 
symmetry on the real line.
\end{enumerate}
\subsubsection{Nonlocal Painlev\'e equations}
It is interesting to note that the shift properties between solutions of the unshifted and shifted nonlocal NLS equations manifest in their self-similar reductions to Painlev\'e type equations as well. In fact, we have the following statements:
\begin{enumerate}
\item The space-shifted nonlocal NLS Eq.~(\ref{shifted-PT-nonloc-NLS})
can be reduced to the nonlocal Painlev\'e equation 
\begin{equation}
    f_{zz}(z)+izf_{z}(z)+(i+\alpha)f(z)-2\sigma f^{2}(z)f^{*}(-z)=0 ,
\end{equation}
via the transformation
\begin{equation}
q(x,t)=\varphi^{1/2}f(z),\;\;\varphi=\frac{e^{i\alpha\log t}}{2t},\;\;z=\frac{x-x_{0}/2}{(2t)^{1/2}} .
\end{equation}
Note that the self-similar variable $z$ includes the shift parameter $x_0$.
\item The time-shifted nonlocal NLS Eq.~(\ref{shifted-t-only-nonloc-NLS})
can be reduced to the nonlocal Painlev\'e equation 
\begin{equation}
    f_{zz}(z)+izf(z)+if(z)-2\sigma\kappa f^{2}(z)f(\kappa z)=0 ,
\end{equation}
where $\kappa$ can be $i$ or $-i$ depending on the choice of branch, using the change of variables
\begin{equation}
    q(x,t)=\frac{1}{(2t-t_{0})^{1/2}}f(z),\;\;z=\frac{x}{(2t-t_{0})^{1/2}} .
\end{equation}
\item The space-time-shifted nonlocal NLS Eq.~(\ref{shifted-x-t-nonloc-NLS})
can be reduced to the nonlocal Painlev\'e equation 
\begin{equation}
    f_{zz}(z)+izf(z)+if(z)-2\sigma\kappa f^{2}(z)f(-\kappa z)=0
\end{equation}
using the change of variables 
\begin{equation}
    q(x,t)=\frac{1}{(2t-t_{0})^{1/2}}f(z),\;\;z=\frac{x-x_{0}/2}{(2t-t_{0})^{1/2}} .
\end{equation}
\item Scaling symmetries: To explain the form of the similarity variables above, we make a note of the pseudo-scaling invariances of these shifted nonlocal evolution equations.
\begin{itemize}
\item Space-shift case: If $q(x,t)$ is a solution of the space-shifted nonlocal NLS Eq.~(\ref{shifted-PT-nonloc-NLS})
then $q_{\lambda}(x,t)\equiv\lambda q(\lambda x,\lambda^{2}t)$ is a solution of 
\begin{equation}
    i\partial_{t}q_{\lambda}(x,t)=\partial_{x}^{2}q_{\lambda}(x,t)-2\sigma q_{\lambda}^{2}(x,t)q_{\lambda}^{*}(\lambda x_{0}-x,t) .
\end{equation}
\item Time-shift case: If $q(x,t)$ is a solution of the time-shifted nonlocal NLS Eq.~(\ref{shifted-t-only-nonloc-NLS})
then $q_{\lambda}(x,t)\equiv\lambda q(\lambda x,\lambda^{2}t)$ is a solution of 
\begin{equation}
    i\partial_{t}q_{\lambda}(x,t)=\partial_{x}^{2}q_{\lambda}(x,t)-2\sigma q_{\lambda}^{2}(x,t)q_{\lambda}(x,\lambda^{2}t_{0}-t) .
\end{equation}
\item Space-time-shift case: If $q(x,t)$ is a solution of the space-time-shifted nonlocal NLS Eq.~(\ref{shifted-x-t-nonloc-NLS})
then $q_{\lambda}(x,t)\equiv\lambda q(\lambda x,\lambda^{2}t)$ is a solution of 
\begin{equation}
    i\partial_{t}q_{\lambda}(x,t)=\partial_{x}^{2}q_{\lambda}(x,t)-2\sigma q_{\lambda}^{2}(x,t)q_{\lambda}(\lambda x_{0}-x,\lambda^{2}t_{0}-t) .
\end{equation}
\end{itemize}
\end{enumerate}
\subsubsection{Relation between inverse scattering and Fourier transforms}
It is a well-known result in Fourier analysis that the Fourier transform of a function with a shifted argument is related to its unshifted counterpart by a phase factor. Let $\mathcal{F}$ denote the Fourier transform defined by 
\begin{equation}
\mathcal{F}[f(x)](k)=\int_{\mathbb{R}}f(x)e^{-ikx}dx, 
\end{equation}
where $f$ is a square-integrable function. Applying this transform to \eqref{connection} we find
\begin{equation}
\label{connection_ft}
  \mathcal{F}[q(x,t)](k)
  =
  e^{ikx_{0}/2}\mathcal{F}[ q_{\text{un}}(x,t)](k).
\end{equation}
As the IST is viewed as a generalization of the classical Fourier transform \cite{AKNS}, one can in fact derive relationships similar to \eqref{connection_ft} between the scattering data obtained for the unshifted nonlocal NLS equations (see for example \cite{AM_Nonlinearity}) 
and the scattering data obtained in the present work for shifted cases. These relations are computed explicitly in Section 3.3.
\subsection{Discrete systems}
The above discussions related to functional connections between solution sets of continuous shifted and non-shifted space-time nonlocal NLS equations persist in the discrete domain. Indeed, consider the Ablowitz-Ladik $(Q_n,R_n)$ system given by
\begin{eqnarray}
\label{discrete_qr1}
i\partial_{t}Q_{n}(t)&=&Q_{n+1}(t)-2Q_{n}(t)+Q_{n-1}(t)-Q_{n}(t)R_{n}(t)(Q_{n+1}(t)+Q_{n-1}(t)),\\
\label{discrete_qr2}
-i\partial_{t}R_{n}(t)&=&R_{n+1}(t)-2R_{n}(t)+R_{n-1}(t)-Q_{n}(t)R_{n}(t)(R_{n+1}(t)+R_{n-1}(t)).
\end{eqnarray}
In analogy to the continuous cases mentioned earlier, the coupled discrete system 
\eqref{discrete_qr1}-\eqref{discrete_qr2} admits several shifted nonlocal symmetry reductions first found in \cite{shift2}. Namely (recall that $\sigma=\pm1$),
\begin{eqnarray}
\label{RQ1}
    R_{n}(t)&=&\sigma Q^{*}_{n_{0}-n}(t),\;n_{0}\in\mathbb{Z},\\
    \label{RQ2}
    R_{n}(t)&=&\sigma Q_{n}(t_{0}-t),\;t_{0}\in\mathbb{C},\\
    \label{RQ3}
    R_{n}(t)&=&\sigma Q_{n_{0}-n}(t_{0}-t),\;n_{0}\in\mathbb{Z}[i],\;t_{0}\in\mathbb{C}.
\end{eqnarray}
In \eqref{RQ3}, the notation $n_{0}\in\mathbb{Z}[i]$ means that the real and imaginary parts of the complex parameter $n_{0}$ are integers.
Customarily, we take $n$ to be an integer throughout, though it is worth noting that the system \eqref{discrete_qr1}-\eqref{discrete_qr2} can be solved by IST for generic $n$ without this restriction to the integers.
 The above reductions lead to the integrable discrete space, time, and space-time NLS equations, respectively given by 
\begin{eqnarray}
\label{discrete_q_eqn1}
i\partial_{t}Q_{n}(t)&=&Q_{n+1}(t)-2Q_{n}(t)+Q_{n-1}(t)
- \sigma Q_{n}(t) Q^{*}_{n_{0}-n}(t) 
\left(  Q_{n+1}(t)+Q_{n-1}(t)  \right),\\
\label{discrete_q_eqn2}
i\partial_{t}Q_{n}(t)&=&Q_{n+1}(t)-2Q_{n}(t)+Q_{n-1}(t)
- \sigma Q_{n}(t) Q_{n}(t_{0}-t) 
\left(  Q_{n+1}(t)+Q_{n-1}(t)  \right),\\
\label{discrete_q_eqn3}
i\partial_{t}Q_{n}(t)&=&Q_{n+1}(t)-2Q_{n}(t)+Q_{n-1}(t)
-\sigma Q_{n}(t) Q_{n_{0}-n}(t_{0}-t) 
\left(  Q_{n+1}(t)+Q_{n-1}(t)  \right).
\end{eqnarray}
Now, let $Q^{\text{un}}_{n}(t)$ denote a solution to the discrete (unshifted) spatially nonlocal equation
\begin{equation}
\label{unshifted_PT_NLS_eqn1a}
i\partial_{t}Q^{\text{un}}_{n}(t)=Q^{\text{un}}_{n+1}(t)-2Q^{\text{un}}_{n}(t) +Q^{\text{un}}_{n-1}(t)
- \sigma Q^{\text{un}}_{n}(t) Q^{\text{un}*}_{-n}(t) 
\left(  Q^{\text{un}}_{n+1}(t) + Q^{\text{un}}_{n-1}(t)  \right). 
\end{equation}
Then,
\begin{equation}
    Q_{n}(t) \equiv Q^{\text{un}}_{n-n_{0}/2}(t),
    \end{equation}
solves the \textit{shifted} evolution equation (\ref{discrete_q_eqn1}). Similar arguments can be made for the time and space-time shifted cases as well. For example, if now $Q^{\text{un}}_{n}(t)$ represents a solution to the (unshifted) space-time discrete nonlocal NLS
\begin{equation}
\label{unshifted_RST_NLS_eqn1a}
i\partial_{t}Q^{\text{un}}_{n}(t)=Q^{\text{un}}_{n+1}(t)-2Q^{\text{un}}_{n}(t) +Q^{\text{un}}_{n-1}(t)
- \sigma Q^{\text{un}}_{n}(t) Q^{\text{un}}_{-n}(-t) 
\left(  Q^{\text{un}}_{n+1}(t) + Q^{\text{un}}_{n-1}(t)  \right), 
\end{equation}
it then follows that,
\begin{equation}
    Q_{n}(t) \equiv Q^{\text{un}}_{n-n_{0}/2}(t - t_0/2),
    \end{equation}
is a solution to the \textit{shifted} problem (\ref{discrete_q_eqn3}). Finally, if the semi-discrete Fourier transform is defined by 
\begin{equation}
\mathcal{F}[f_{n}](Z)
=
\sum_{n\in\mathbb{Z}}f_{n}Z^{n} ,\;\; | Z| =1,
\end{equation}
 then we have the following result
\begin{equation}
    \label{disc-ft-connection}
    \mathcal{F}[Q_{n}](Z)=\mathcal{F}[Q^{\text{un}}_{n-n_{0}/2}](Z)=Z^{n_{0}/2}
    \mathcal{F}[Q^{\text{un}}_{n}](Z).
\end{equation}
This is analogous to the continuous result given in \eqref{connection_ft}. In section 8.3, we find a similar relationship between the scattering data associated with the shifted and unshifted Ablowitz-Ladik scattering problems.
\section{Direct scattering -- Continuous case}
\label{direct_scattering_cont}
\subsection{Linear pair}
We begin with the AKNS spectral problem \cite{AKNS}
\begin{eqnarray}
    \label{scat}
    v_{x}&=&\mathcal{X}v,\;\;\mathcal{X}=\begin{pmatrix}-ik&q(x,t)\\r(x,t)&ik\end{pmatrix},\\
    \label{time}
    v_{t}&=&\mathcal{T}v.
\end{eqnarray}
where $v(x,t,k)=(v_{1}(x,t,k),v_{2}(x,t,k))^{T}$ with $T$ denoting matrix transpose, and $q(x,t)$ and $r(x,t)$ 
are complex-valued functions of the real variables $x,t\in\mathbb{R}.$
Here, $k$ is a complex spectral parameter independent of $x,t$. Furthermore, $q$ and $r$ are referred to as potentials in the AKNS scattering problem and are assumed throughout the rest of the paper to decay rapidly to zero as $x\rightarrow\pm\infty$; 
for the scattering problem we are studying (\ref{scat}), AKNS \cite{AKNS} established that $q,r \in L^1$ is sufficient.  
If $\mathcal{T}$ is chosen to be of the form
\begin{equation}
    \label{time-nls}
    \mathcal{T}_{\text{NLS}}=\begin{pmatrix}2ik^{2}+iqr&
2kq-iq_{x}\\
2kr+ir_{x}&-2ik^{2}-iqr\end{pmatrix},
\end{equation}
then the compatibility condition between \eqref{scat} and \eqref{time} (namely, $v_{xt}=v_{tx}$) results in the coupled NLS system given in \eqref{q-sys}-\eqref{r-sys}. Alternatively, if $\mathcal{T}$ is given by
\begin{equation}
    \label{time-mkdv}
    \mathcal{T}_{\text{mKdV}}=\begin{pmatrix}
        -4ik^{3}-2ikqr+rq_{x}-qr_{x}&4k^{2}q+2ikq_{x}+2q^{2}r-q_{xx}\\4k^{2}r-2ikr_{x}+2qr^{2}-r_{xx}&4ik^{3}+2ikqr-rq_{x}+qr_{x}
    \end{pmatrix},
\end{equation}
then the compatibility condition gives rise to the mKdV system 
\eqref{q-sys-mkdv}-\eqref{r-sys-mkdv}.

\subsection{Eigenfunctions and scattering data}

We now summarize the major components needed to carry out the inverse scattering transform for generic potentials. The first step involves solving the direct scattering problem \eqref{scat} at the initial instant. For local in time reductions, the initial instant can be any time (without loss of generality we take it to be $t=0$). However, it is important to note that for the time-shifted reductions, 
the initial instant is set to be $t=t_{0}/2$ so that the direct problem is well-posed, i.e. we assume an initial condition $q(x,t_{0}/2)$ is provided. 
\\\\
\noindent Consider the following eigenfunctions of \eqref{scat} defined by the boundary conditions they satisfy:
\begin{eqnarray}
\label{eigenphi}
&\phi(x,t,k)\sim e^{-ikx}\mathbf{\hat{e}_{1}},\;\;\;\;&\bar{\phi}(x,t,k)\sim e^{ikx}\mathbf{\hat{e}_{2}},\;\;\;\;\;\;\;\text{as}\;x\rightarrow-\infty,\\
\label{eigenpsi}
&\psi(x,t,k)\sim e^{ikx}\mathbf{\hat{e}_{2}},\;\;\;\;&\bar\psi(x,t,k)\sim e^{-ikx}\mathbf{\hat{e}_{1}},\;\;\;\;\text{as}\;x\rightarrow+\infty.
\end{eqnarray}
Here, and throughout the rest of the paper, $\mathbf{\hat{e}_{1}}=(1,0)^{T}$ and $\mathbf{\hat{e}_{2}}=(0,1)^{T}$. Next, since each of the eigenfunction pairs $\{\phi,\bar\phi\}$ and $\{\psi,\bar\psi\}$ are separately linearly independent, we may construct an expansion of the form
\begin{eqnarray}
\label{phiexp}
\phi(x,t,k)=a(t,k)\bar\psi(x,t,k)+b(t,k)\psi(x,t,k),\\
\label{barphiexp}
\bar\phi(x,t,k)=\bar{a}(t,k)\psi(x,t,k)+\bar{b}(t,k)\bar\psi(x,t,k),
\end{eqnarray}
where $a,b,\bar{a},\bar{b}$ are referred to as the scattering data and are given by
\begin{eqnarray}
\label{awron}
a(t,k)&=&W(\phi(x,t,k),\psi(x,t,k)),\\
\label{barawron}
\bar{a}(t,k)&=&W(\bar\psi(x,t,k),\bar\phi(x,t,k)),\\
\label{bwron}
b(t,k)&=&W(\bar\psi(x,t,k),\phi(x,t,k)),\\
\label{barbwron}
\bar{b}(t,k)&=&W(\bar\phi(x,t,k),\psi(x,t,k)),
\end{eqnarray}
with $W$ being the Wronksian defined by $W(u,v)=u_{1}v_{2}-v_{1}u_{2}$ with $u=(u_{1},u_{2})^{T}$ and $v=(v_{1},v_{2})^{T}$.
Furthermore, it can be shown that the scattering data satisfy
\begin{equation}
    a(t,k)\bar{a}(t,k)-b(t,k)\bar{b}(t,k)=1.
\end{equation}
It will also be useful to define the following modified eigenfunctions that satisfy constant boundary conditions:
\begin{eqnarray}
\label{jostm}
&M(x,t,k)=e^{ikx}\phi(x,t,k),\;\;&\bar{M}(x,t,k)=e^{-ikx}\bar\phi(x,t,k),\\
\label{jostn}
&N(x,t,k)=e^{-ikx}\psi(x,t,k),\;\;&\bar{N}(x,t,k)=e^{ikx}\bar\psi(x,t,k).
\end{eqnarray}
Importantly, it can be shown that $M$ and $N$ can be analytically continued into the upper half complex $k$-plane, while $\bar M$ and $\bar N$ are analytic in the lower half complex $k$-plane. These analyticity properties will be crucial in the formulation of the inverse scattering problem. Particularly, one can rewrite \eqref{phiexp} and \eqref{barphiexp} as
\begin{eqnarray}
\label{rh1}
    \mu(x,t,k)&=&\bar{N}(x,t,k)+\rho(t,k)e^{2ikx}N(x,t,k),\\
    \label{rh2}
    \bar\mu(x,t,k)&=&N(x,t,k)+\bar\rho(t,k)e^{-2ikx}\bar{N}(x,t,k),
\end{eqnarray}
where $\mu=M/a$ and $\bar\mu=\bar{M}/\bar{a}$, and the so-called reflection coefficients are given by $\rho=b/a$ and $\bar\rho=\bar{b}/\bar{a}$. The system \eqref{rh1}-\eqref{rh2} will later be viewed as a Riemann-Hilbert problem on the real $k$-axis.
\subsection{Relation between the scattering data for the spatially shifted and unshifted AKNS systems}
Here, we make a note of the scattering-space relationships between spatially shifted nonlocal equations and their unshifted counterparts, as mentioned in Section 2. Since the relations here apply to all reductions with a spatial shift, we suppress the time argument throughout this section. First, recall the physical-space relationship \eqref{connection} $q(x)=q_{\text{un}}(x-x_0/2)\equiv q_{\text{un}}(y)$, where $q_{\text{un}}(y)$ solves the \textit{unshifted} problem (\ref{unshifted-PT-nonloc-NLS}). 
Let  $\Phi,\bar\Phi,\Psi,\bar\Psi$ denote the eigenfunctions associated with the unshifted problem, defined by
\begin{eqnarray}
\label{eigenphi-un}
&\Phi(y,k)\sim e^{-iky}\mathbf{\hat{e}_{1}},\;\;\;\;&\bar{\Phi}(y,k)\sim e^{iky}\mathbf{\hat{e}_{2}},\;\;\;\;\;\;\;\text{as}\;y\rightarrow-\infty,\\
\label{eigenpsi-un}
&\Psi(y,k)\sim e^{iky}\mathbf{\hat{e}_{2}},\;\;\;\;&\bar\Psi(y,k)\sim e^{-iky}\mathbf{\hat{e}_{1}},\;\;\;\;\text{as}\;y\rightarrow+\infty.
\end{eqnarray}
Since $y=x-x_{0}/2$, comparing with \eqref{eigenphi}-\eqref{eigenpsi} gives the relationships
\begin{eqnarray}
\label{unshift1}
    &\phi(x,k)=e^{-ikx_{0}/2}\Phi(x-x_{0}/2,k),\;\;\;\;&\bar\phi(x,k)=e^{ikx_{0}/2}\bar\Phi(x-x_{0}/2,k),\\
    \label{unshift2}
     &\psi(x,k)=e^{ikx_{0}/2}\Psi(x-x_{0}/2,k),\;\;\;\;&\bar\psi(x,k)=e^{-ikx_{0}/2}\bar\Psi(x-x_{0}/2,k).
\end{eqnarray}
Furthermore, let $A,B,\bar{A},\bar{B}$ denote the scattering data associated with the unshifted problem. The Wronskian relations \eqref{awron}-\eqref{barbwron} together with \eqref{unshift1}-\eqref{unshift2} lead to the results
\begin{eqnarray}
\label{unshifta}
    &a(k)=A(k),\;\;\;\;&\bar{a}(k)=\bar{A}(k),\\
    \label{unshiftb}
    &b(k)=e^{-ikx_{0}}B(k),\;\;\;\;&\bar{b}(k)=e^{ikx_{0}}\bar B(k).
\end{eqnarray}
Note the similarity between the property of the Fourier transform given in \eqref{connection_ft} and the relations in \eqref{unshiftb}.
\section{Symmetries in scattering space -- Continuous case}
In this section, we use the various shifted nonlocal symmetry reductions to derive the corresponding symmetry relationships between the eigenfunctions and scattering data. Throughout this section, for the sake of convenience we will make frequent use of the following definition
\begin{equation}
    \Lambda_{\pm}\equiv\begin{pmatrix}0&\pm\sigma\\1&0\end{pmatrix},\;\;\Lambda_{\pm}^{-1}\equiv\begin{pmatrix}0&1\\\pm\sigma&0\end{pmatrix}.
\end{equation}

\subsection{Time shifted symmetry reduction: $r(x,t)=\sigma q(x,t_{0}-t)$}
\subsubsection{Symmetries between the eigenfunctions}
Let $v(x,t,k)=(v_{1}(x,t,k),v_{2}(x,t,k))^{T}$ be any solution to the system \eqref{scat} with $r(x,t)=\sigma q(x,t_{0}-t)$.
Letting $t\rightarrow t_{0}-t$ and $k\rightarrow-k$ in \eqref{scat} we find
\begin{eqnarray}
\sigma v_{2x}(x,t_{0}-t,-k)&=&-ik\sigma v_{2}(x,t_{0}-t,-k)+q(x,t)v_{1}(x,t_{0}-t,-k),\\
v_{1x}(x,t_{0}-t,-k)&=&ik v_{1}(x,t_{0}-t,-k)+\sigma q(x,t_{0}-t)\sigma v_{2}(x,t_{0}-t,-k).\nonumber
\end{eqnarray}
Therefore, this leads to the symmetry relationship:
\begin{equation*}
{\rm If}\; v(x,t,k)\;
{\rm solves}\; (\ref{scat})\; {\rm with}\; r(x,t)=\sigma q(x,t_{0}-t) \; {\rm so \; does}\;
\Lambda_{+}v(x,t_{0}-t,-k) .
\end{equation*}
Next, let $\phi(x,t,k)$ be the solution of \eqref{scat} satisfying the boundary condition given in the left part of \eqref{eigenphi}. From the above statement, we know that $\Lambda_{+}\phi(x,t_{0}-t,-k)$ is also a solution. Additionally, suppose $\bar\phi(x,t,k)$ is the solution of \eqref{scat} satisfying the boundary condition given in the right part of \eqref{eigenphi}. Comparing respective boundary conditions, one finds the symmetry relationship:
\begin{equation}
    \label{phi-sym-ts}
    \bar\phi(x,t,k)=\Lambda_{+}\phi(x,t_{0}-t,-k).
\end{equation}
Similarly, let $\psi(x,t,k)$ and $\bar\psi(x,t,k)$ satisfy the boundary conditions specified in \eqref{eigenpsi}. Thus, we find:
\begin{equation}
    \label{psi-sym-ts}
    \bar{\psi}(x,t,k)=\Lambda_{+}^{-1}\psi(x,t_{0}-t,-k).
\end{equation}
The symmetries \eqref{phi-sym-ts} and \eqref{psi-sym-ts} induce the following relations between the modified eigenfunctions defined in \eqref{jostm}-\eqref{jostn}:
\begin{eqnarray}
    \label{m-sym-ts}
    \bar M(x,t,k)&=&\Lambda_{+}M(x,t_{0}-t,-k),\\
    \label{n-sym-ts}
    \bar N(x,t,k)&=&\Lambda_{+}^{-1}N(x,t_{0}-t,-k).
\end{eqnarray}

\subsubsection{Symmetries of the scattering data}
To derive a relationship between the scattering data $a$ and $\bar{a}$, we begin with \eqref{barawron}. After applying the previously derived symmetries between the eigenfunctions we find
\begin{equation}
\bar{a}(t,k)={\psi}_{2}(x,t_{0}-t,-k)\phi_{1}(x,t_{0}-t,-k)-\sigma\psi_{1}(x,t_{0}-t,-k)\sigma\phi_{2}(x,t_{0}-t,-k).
\end{equation}
However, from \eqref{awron} we also have
\begin{equation}
a(t_{0}-t,-k)=\phi_{1}(x,t_{0}-t,-k)\psi_{2}(x,t_{0}-t,-k)-\phi_{2}(x,t_{0}-t,-k)\psi_{1}(x,t_{0}-t,-k).
\end{equation}
Comparing these two results, one can see that
\begin{equation}
    \label{a-sym-ts}
    \bar{a}(t,k)=a(t_{0}-t,-k).
\end{equation}
We will find later that $a$ and $\bar{a}$ are actually time independent, as such there is a relationship between the zeros of these functions:
\begin{equation}
\label{eig-sym-cont}
\text{If $a(k_{j})=0$ and $\bar{k}_j=-k_{j}$, then $\bar{a}(\bar{k}_{j})=0$.}
\end{equation}
To find a symmetry relationship between $b$ and $\bar{b}$, we start from \eqref{barbwron} and apply \eqref{phi-sym-ts}-\eqref{psi-sym-ts} to get
\begin{equation}
\label{bbar}
\bar{b}(t,k)=\sigma\phi_{2}(x,t_{0}-t,-k)\bar\psi_{1}(x,t_{0}-t,-k)-\phi_{1}(x,t_{0}-t,-k)\sigma\bar\psi_{2}(x,t_{0}-t,-k).
\end{equation}
On the other hand, from \eqref{bwron} we have
\begin{equation}
\label{b}
b(t_{0}-t,-k)=\bar\psi_{1}(x,t_{0}-t,-k)\phi_{2}(x,t_{0}-t,-k)-\bar\psi_{2}(x,t_{0}-t,-k)\phi_{1}(x,t_{0}-t,-k).
\end{equation}
From (\ref{bbar}) and (\ref{b}) we arrive at the symmetry condition
\begin{equation}
    \label{b-sym-ts}
    \bar{b}(t,k)=\sigma b(t_{0}-t,-k).
\end{equation}

\subsection{Space-time shift reduction: $r(x,t)=\sigma q(x_{0}-x,t_{0}-t)$}
\subsubsection{Symmetries between the eigenfunctions}
Assume that $v(x,t,k)$ is any solution of \eqref{scat} with $r(x,t)=\sigma q(x_{0}-x,t_{0}-t)$.
Performing the change of variables $x\rightarrow x_{0}-x$ and $t\rightarrow t_{0}-t$ and rearranging 
terms gives
\begin{eqnarray}
\sigma v_{1x}(x_{0}-x,t_{0}-t,k)&=&ik\sigma v_{1}(x_{0}-x,t_{0}-t,k)-\sigma q(x_{0}-x,t_{0}-t)v_{2}(x_{0}-x,t_{0}-t,k),\\
-v_{2x}(x_{0}-x,t_{0}-t,k)&=&ikv_{2}(x_{0}-x,t_{0}-t,k)+\sigma q(x,t)v_{1}(x_{0}-x,t_{0}-t,k).\nonumber
\end{eqnarray}
From this we can deduce the following:
\begin{equation*}
{\rm If}\; v(x,t,k)\;
{\rm solves}\; (\ref{scat})\; {\rm with}\; r(x,t)=\sigma q(x_{0}-x,t_{0}-t) \; {\rm so \; does}\;
\Lambda_{-}v(x_{0}-x,t_{0}-t,k) .
\end{equation*}
Since $\phi(x,t,k)$ (as defined in \eqref{eigenphi}) is a solution of \eqref{scat}, we know that $\Lambda_{-}\phi(x_{0}-x,t_{0}-t,k)$ is also a solution. Comparing boundary conditions, it can be found that:
\begin{equation}
\label{nobar-sym-sts}
\psi(x,t,k)=e^{ikx_{0}}\Lambda_{-}\phi(x_{0}-x,t_{0}-t,k).
\end{equation}
Note that in contrast to the time shifted case, \eqref{nobar-sym-sts} provides a connection between the two eigenfunctions defined in the upper half plane. A similar relationship can be derived for the eigenfunctions defined in the lower half plane. Indeed, one finds:
\begin{equation}
\label{bar-sym-sts}
\bar\psi(x,t,k)=e^{-ikx_{0}}\Lambda_{-}^{-1}\bar\phi(x_{0}-x,t_{0}-t,k).
\end{equation}
These symmetries in turn produce corresponding symmetries between the modified eigenfunctions \eqref{jostm}-\eqref{jostn}, namely:
\begin{eqnarray}
    \label{nobar-jost-sym-sts}
    N(x,t,k)&=&\Lambda_{-}M(x_{0}-x,t_{0}-t,k),\\
    \label{bar-jost-sym-sts}
    \bar N(x,t,k)&=&\Lambda_{-}^{-1}\bar M(x_{0}-x,t_{0}-t,k).
\end{eqnarray}

\subsubsection{Symmetries of the scattering data}
Contrary to the time-shifted case, it turns out that the scattering data $a$ and $\bar{a}$ are not related through a symmetry condition. Instead, observe that from \eqref{awron}, after applying the symmetry \eqref{nobar-sym-sts} one finds
\begin{equation}
a(t,k)=e^{ikx_{0}}[\phi_{1}(x,t,k)\phi_{1}(x_{0}-x,t_{0}-t,k)+\sigma\phi_{2}(x,t,k)\phi_{2}(x_{0}-x,t_{0}-t,k)].
\end{equation}
Letting $x\rightarrow x_{0}-x$ and $t\rightarrow t_{0}-t$ shows that $a$ has the property
\begin{equation}
    \label{a-sym-sts}
    a(t,k)=a(t_{0}-t,k).
\end{equation}
Furthermore, using a similar derivation, one can show that
\begin{equation}
    \label{bara-sym-sts}
    \bar a(t,k)=\bar a(t_{0}-t,k).
\end{equation}
As we shall note later, $a$ and $\bar{a}$ are in fact time independent, in which case the previous two results become trivial.
Next, starting from \eqref{barbwron} and making use of \eqref{nobar-sym-sts},
\begin{equation}
\bar{b}(t,k)=e^{ikx_{0}}[\bar\phi_{1}(x,t,k)\phi_{1}(x_{0}-x,t_{0}-t,k)+\sigma\bar\phi_{2}(x,t,k)\phi_{2}(x_{0}-x,t_{0}-t,k)].
\end{equation}
On the other hand, starting from \eqref{bwron} and using \eqref{bar-sym-sts},
\begin{equation}
\label{b-wron-with-sym}
b(t,k)=e^{-ikx_{0}}[\bar\phi_{2}(x_{0}-x,t_{0}-t,k)\phi_{2}(x,t,k)+\sigma\bar\phi_{1}(x_{0}-x,t_{0}-t,k)\phi_{1}(x,t,k)].
\end{equation}
Letting $x\rightarrow x_{0}-x$ and $t\rightarrow t_{0}-t$ in \eqref{b-wron-with-sym}, we find
\begin{equation}
    \label{b-sym-sts}
    \bar b(t,k)=\sigma e^{2ikx_{0}}{b}(t_{0}-t,k).
\end{equation}

\subsection{Space shift reduction: $r(x,t)=\sigma q^{*}(x_{0}-x,t)$}
\subsubsection{Symmetries between the eigenfunctions}
Assume that $v(x,t,k)$ is any solution of \eqref{scat} with $r(x,t)=\sigma q^{*}(x_{0}-x,t)$.
Letting $x\rightarrow x_{0}-x$ and $k\rightarrow-k^{*}$, taking the complex conjugate, and rearranging gives
\begin{eqnarray}
\sigma v_{1x}^{*}(x_{0}-x,t,-k^{*})&=&ik\sigma v_{1}^{*}(x_{0}-x,t,-k^{*})-\sigma q^{*}(x_{0}-x,t)v_{2}^{*}(x_{0}-x,t,-k^{*}),\\
-v_{2x}^{*}(x_{0}-x,t,-k^{*})&=&ikv_{2}^{*}(x_{0}-x,t,-k^{*})+\sigma q(x,t)v_{1}^{*}(x_{0}-x,t,-k^{*}),\nonumber
\end{eqnarray}
from which we can deduce the following:
\begin{equation*}
{\rm If}\; v(x,t,k)\;
{\rm solves}\; (\ref{scat})\; {\rm with}\; r(x,t)=\sigma q^{*}(x_{0}-x,t) \; {\rm so \; does}\;
\Lambda_{-}v^{*}(x_{0}-x,t,-k^{*}) .
\end{equation*}
In a similar fashion to the space-time shifted case, by comparing boundary conditions one can obtain the following symmetries:
\begin{eqnarray}
    \label{nobar-sym-ss}
    \psi(x,t,k)&=&e^{ikx_{0}}\Lambda_{-}\phi^{*}(x_{0}-x,t,-k^{*}),\\
    \label{bar-sym-ss}
    \bar\psi(x,t,k)&=&e^{-ikx_{0}}\Lambda_{-}^{-1}\bar\phi^{*}(x_{0}-x,t,-k^{*}),
\end{eqnarray}
which in turn lead to symmetries between the modified eigenfunctions:
\begin{eqnarray}
    \label{nobar-jost-sym-ss}
    N(x,t,k)&=&\Lambda_{-}M^{*}(x_{0}-x,t,-k^{*}),\\
       \label{bar-jost-sym-ss}
    \bar N(x,t,k)&=&\Lambda_{-}^{-1}\bar M^{*}(x_{0}-x,t,-k^{*}).
\end{eqnarray}

\subsubsection{Symmetries of the scattering data}
Starting from \eqref{awron} and using \eqref{nobar-sym-ss}, we have
\begin{equation}
a(t,k)=e^{ikx_{0}}[\phi_{1}(x,t,k)\phi_{1}^{*}(x_{0}-x,t,-k^{*})+\sigma\phi_{2}(x,t,k)\phi_{2}^{*}(x_{0}-x,t,-k^{*})].
\end{equation}
Letting $x\rightarrow x_{0}-x$, $k\rightarrow-k^{*}$, and complex conjugating leads to the result:
\begin{equation}
    \label{a-sym-ss}
    a(t,k)=a^{*}(t,-k^{*}).
\end{equation}
Also, the scattering data $\bar{a}$ satisfies an analogous property:
\begin{equation}
\label{bara-sym-ss}
    \bar a(t,k)=\bar a^{*}(t,-k^{*}).
\end{equation}
Note that the previous results imply the following fact:
\begin{align*}
\label{eig-PT-conclusion-cont}
&
 \text{If}~ a(k_{j})=0~\text{and}~ \bar{a}(\bar{k}_{j})=0,~\text{then}~ a(-k_{j}^{*})=0~\text{and}~\bar{a}(-\bar{k}_{j}^{*})=0 .
\end{align*}
Next, start form \eqref{barbwron} and apply \eqref{nobar-sym-ss} to obtain
\begin{equation}
\bar{b}(t,k)=e^{ikx_{0}}[\bar\phi_{1}(x,t,k)\phi_{1}^{*}(x_{0}-x,t,-k^{*})+\sigma\bar\phi_{2}(x,t,k)\phi_{2}^{*}(x_{0}-x,t,-k^{*})].
\end{equation}
Additionally, from \eqref{bwron} using \eqref{bar-sym-ss} we have
\begin{equation}
\label{b-wron-with-sym-ss}
b(t,k)=e^{-ikx_{0}}[\bar\phi_{2}^{*}(x_{0}-x,t,-k^{*})\phi_{2}(x,t,k)+\sigma\bar\phi_{1}^{*}(x_{0}-x,t,-k^{*})\phi_{1}(x,t,k)].
\end{equation}
Let $x\rightarrow x_{0}-x$, conjugate, and let $k\rightarrow-k$ (noting that $b$ is only defined for real $k$), to get
\begin{equation}
\label{b-sym-ss}
\bar{b}(t,k)=\sigma e^{2ikx_{0}}b^{*}(t,-k^{}) \;, \;\; k\in\mathbb{R}
\end{equation}
\subsection{Conjugate space-time shifted reduction (mKdV)}
\subsubsection{Symmetries between the eigenfunctions}
Assume that $v(x,t,k)$ is any solution of \eqref{scat} with $r(x,t)=\sigma q^{*}(x_{0}-x,t_{0}-t)$. Letting $x\rightarrow x_{0}-x$, $t\rightarrow t_{0}-t$, $k\rightarrow-k^{*}$; taking the complex conjugate; and rearranging gives
\begin{eqnarray}
\sigma v_{1x}^{*}(x_{0}-x,t_{0}-t,-k^{*})&=&ik\sigma v_{1}^{*}(x_{0}-x,t_{0}-t,-k^{*})-\sigma q^{*}(x_{0}-x,t_{0}-t)v_{2}^{*}(x_{0}-x,t_{0}-t,-k^{*}), \nonumber \\
-v_{2x}^{*}(x_{0}-x,t_{0}-t,-k^{*})&=&ikv_{2}^{*}(x_{0}-x,t_{0}-t,-k^{*})+\sigma q(x,t)v_{1}^{*}(x_{0}-x,t_{0}-t,-k^{*}),
\end{eqnarray}
from which the we obtain the following conclusion:
\begin{equation*}
{\rm If}\; v(x,t,k)\;
{\rm solves}\; (\ref{scat})\; {\rm with}\; r(x,t)=\sigma q^{*}(x_{0}-x,t_{0}-t) \; {\rm so \; does}\;
\Lambda_{-}v^{*}(x_{0}-x,t_{0}-t,-k^{*}) .
\end{equation*}
In a similar fashion to the space-time and space shifted reductions, by comparing boundary conditions one can obtain the following symmetries:
\begin{eqnarray}
    \label{nobar-sym-csts}
    \psi(x,t,k)&=&e^{ikx_{0}}\Lambda_{-}\phi^{*}(x_{0}-x,t_{0}-t,-k^{*}),\\
    \label{bar-sym-csts}
    \bar\psi(x,t,k)&=&e^{-ikx_{0}}\Lambda_{-}^{-1}\bar\phi^{*}(x_{0}-x,t_{0}-t,-k^{*}).
\end{eqnarray}
These in turn lead to symmetries between the modified eigenfunctions:
\begin{eqnarray}
    \label{nobar-jost-sym-csts}
    N(x,t,k)=\Lambda_{-}M^{*}(x_{0}-x,t_{0}-t,-k^{*}),\\
       \label{bar-jost-sym-csts}
    \bar N(x,t,k)=\Lambda_{-}^{-1}\bar M^{*}(x_{0}-x,t_{0}-t,-k^{*}).
\end{eqnarray}
\subsubsection{Symmetries of the scattering data}

Since the calculations are very similar to the previous two reductions, we only summarize the results:
\begin{eqnarray}
\label{a-sym-csts}
&&a(t,k)=a^{*}(t_{0}-t,-k^{*}),\;\;\;\bar a(t,k)=\bar a^{*}(t_{0}-t,-k^{*}).\\
\label{eig-PT-conclusion-cont-csts}
 &&\text{If}~ a(k_{j})=0~\text{and}~ \bar{a}(\bar{k}_{j})=0,~\text{then}~ a(-k_{j}^{*})=0~\text{and}~\bar{a}(-\bar{k}_{j}^{*})=0 .\nonumber\\
    \label{b-sym-csts}
    &&\bar{b}(t,k)=\sigma e^{2ikx_{0}}b^{*}(t_{0}-t,-k^{})\;,\;\; k\in\mathbb{R}.
\end{eqnarray}

\section{Inverse scattering and time dependence -- Continuous case}

\subsection{Inverse scattering: Riemann-Hilbert approach}
The next step of the inverse scattering transform is to solve an inverse problem with the goal of constructing an explicit formula for the potentials $q(x,t)$ and $r(x,t)$ using the known scattering data. This is accomplished by reformulating the AKNS scattering problem as a Riemann-Hilbert problem. 
The scattering problem \eqref{scat} can admit discrete eigenvalues. We will assume that these occur at simple zeros of $a(k)$ and $\bar{a}(k)$ in the upper and lower half planes respectively, and will denote them by $\{k_{j}:\Im k_{j}>0\}_{j=1}^{J}$, and $\{\bar k_{j}:\Im \bar k_{j}<0\}_{j=1}^{\bar{J}}$, respectively. Furthermore, it is useful to define the so-called norming constants,
\begin{equation}
    \label{norming-cont}
    C_{j}\equiv\frac{b_{j}}{a_{k}(k_{j})},\;\;\bar{C}_{j}\equiv\frac{\bar{b}_{j}}{\bar{a}_{k}(\bar{k}_{j})},
\end{equation}
where $b_{j}, \bar{b}_{j}$ are bound state coefficients and $a_k$ denotes derivative of $a$ with respect to $k$. The details of the Riemann-Hilbert problem associated with the AKNS scattering problem are discussed thoroughly in \cite{APT_Book}. Presently, we quote the main result of the inverse problem. That is, from \eqref{rh1} and \eqref{rh2} it can be shown that the modified eigenfunctions satisfy a system of algebraic-integral equations involving the discrete eigenvalues, norming constants, and reflection coefficients:
\begin{eqnarray}
\label{N-Nbar-sys-1}
\bar{N}(x,\bar k_{j})&=&\mathbf{\hat{e}_{1}}+\sum_{\ell=1}^{J}\frac{C_{\ell}e^{2ik_{\ell}x}N(x,k_{\ell})}{\bar k_{j}-k_{\ell}}+\frac{1}{2\pi i}\int_{\mathbb{R}}\frac{\rho(\xi)e^{2i\xi x}N(x,\xi)}{\xi-(\bar k_{j}-i0)}d\xi,\\
\label{N-Nbar-sys-2}
N(x,k_{j})&=&\mathbf{\hat{e}_{2}}+\sum_{\ell=1}^{\bar{J}}\frac{\bar{C}_{\ell}e^{-2i\bar{k}_{\ell}x}\bar{N}(x,\bar{k}_{\ell})}{k_{j}-\bar{k}_{\ell}}-\frac{1}{2\pi i}\int_{\mathbb{R}}\frac{\bar\rho(\xi)e^{-2i\xi x}\bar{N}(x,\xi)}{\xi-(k_{j}+i0)}d\xi.
\end{eqnarray}
Once the eigenfunctions are determined from the above system, the potentials can be recovered using the following formulae derived using the asymptotic expansions of the modified eigenfunctions (see \cite{APT_Book} for details):
\begin{eqnarray}
\label{r-recon}
r(x)&=&-2i\sum_{j=1}^{J}e^{2ik_{j}x}C_{j}(t)N_{2}(x,k_{j})+\frac{1}{\pi}\int_{\mathbb{R}}\rho(\xi)e^{2i\xi x}N_{2}(x,\xi)d\xi,\\
\label{q-recon}
q(x)&=&2i\sum_{j=1}^{\bar{J}}e^{-2i\bar{k}_{j}x}\bar{C}_{j}\bar{N}_{1}(x,\bar{k}_{j})+\frac{1}{\pi}\int_{\mathbb{R}}\bar{\rho}(\xi)e^{-2i\xi x}\bar{N}_{1}(x,\xi)d\xi.
\end{eqnarray}

\subsection{Time evolution} 
In this section, we provide the time evolution of all scattering data and norming constants. The time evolution of the scattering data is the primary difference between the inverse scattering transforms associated with the coupled NLS \eqref{q-sys}-\eqref{r-sys} and mKdV \eqref{q-sys-mkdv}-\eqref{r-sys-mkdv} systems due to the difference in the choice of $\mathcal{T}$ in \eqref{time}. 

\subsubsection{NLS time evolution}
If the time-evolution component of the linear pair is chosen to be $\mathcal{T}_{\text{NLS}}$ as in \eqref{time-nls}, the scattering data $a,b$ evolve in time according to the following results (see for example \cite{APT_Book}):
\begin{equation}
\label{time-evolv-a-and-b-explicit_cont}
a_{t}( t,k ) = 0\;,\;\;\;\;\;\;\;\;\;\; 
b_{t}( t ,k) = -4i k^{2}b(t,k)\;.
\end{equation}
Similarly, the other set of scattering data $\bar{a},\bar{b}$ satisfy 
\begin{equation}
\label{time-evolv-abar-and-bbar-explicit_cont}
\bar{a}_{t}( t,k ) =0\;,\;\;\;\;\;
\bar{b}_{t}( t,k ) =  4ik^{2}\bar{b}(t,k)\;.
\end{equation}
For the local in time reductions (in the present work, only the space-shifted case), we have
\begin{eqnarray}
    &a(t,k)=a(0,k),\;\;\;&\bar{a}(t,k)=\bar{a}(0,k),\\
    &b(t,k)=e^{-4ik^{2}t}b(0,k),\;\;\;&\bar{b}(t,k)=e^{4ik^{2}t}\bar{b}(0,k).
\end{eqnarray}
As noted in section III, for all time-shifted reductions, the initial condition must be provided at $t=t_{0}/2$, in which case the initial scattering data will be obtained at that time. As such, we denote the time evolution of the scattering data by
\begin{eqnarray}
    &a(t,k)=a(t_{0}/2,k),\;\;\;&\bar{a}(t,k)=\bar{a}(t_{0}/2,k),\\
    &b(t,k)=e^{-4ik^{2}(t-t_{0}/2)}b(t_{0}/2,k),\;\;\;&\bar{b}(t,k)=e^{4ik^{2}(t-t_{0}/2)}\bar{b}(t_{0}/2,k).
\end{eqnarray}
The evolution of the norming constants $C_j$ and $\bar{C}_j$ defined in equation \eqref{norming-cont}
is given for time-local cases by
 \begin{equation}
\label{C-ell-evolve_cont}
C_j(t)  
 = 
C_j (0) e^{-4ik_{j}^{2} t},\;\;\; 
\bar{C}_j (t)
 =    \bar{C}_j (0) e^{4i\bar{k}_{j}^{2} t},
\end{equation}
and for time-shifted cases by
 \begin{equation}
\label{C-ell-evolve_cont-shifted}
C_j(t)  
 = 
C_j (t_{0}/2) e^{-4ik_{j}^{2} (t-t_{0}/2)},\;\;\; 
\bar{C}_j (t)
 =    \bar{C}_j (t_{0}/2) e^{4i\bar{k}_{j}^{2} (t-t_{0}/2)}.
\end{equation}

\subsubsection{mKdV time evolution}
If the time-evolution component of the linear pair is chosen to be $\mathcal{T}_{\text{mKdV}}$ as in \eqref{time-mkdv}, the scattering data $a$ and $\bar{a}$ are again independent of time, and $b$ and $\bar{b}$ evolve according to
\begin{equation}
\label{time-evol-mkdv}
    b( t,k ) = e^{8ik^{3}(t-t_{0}/2)}b(t_{0}/2,k)\;,\;\;\;\;\;\;\;\;\;\; 
\bar b( t ,k) = e^{-8i k^{3} (t-t_{0}/2)} b(t_{0}/2,k)\;,
\end{equation}
since we are only considering time-shifted reductions to the mKdV system. Following from this, the time evolution of the norming constants is given by
 \begin{equation}
\label{C-ell-evolve_cont_mkdv}
C_j(t)  
 = 
C_j (t_{0}/2) e^{8ik_{j}^{3} (t-t_{0}/2)},\;\;\; 
\bar{C}_j (t)
 =     \bar{C}_j (t_{0}/2) e^{-8i\bar{k}_{j}^{3} (t-t_{0}/2)}.
\end{equation}

\section{Trace formulae and symmetries for $b_j, \bar{b}_j$ -- Continuous case}

For the time-shift reduction case (as well as the classical NLS) the symmetries connect the scattering data and norming 
constants in the upper half plane to their corresponding quantities defined in the lower half plane. This implies that the eigenvalues
$k_j$ and norming constants $C_j$ are counted as free parameters and the values of $\bar{k}_j$ and $\bar{C}_j$ are uniquely determined by 
the underlying symmetries. However, in the cases involving spatial nonlocality, the symmetries of the scattering data and norming constants do {\it not} relate their respective values in the upper and lower half planes. In order to understand the underlying symmetries of the norming constants 
$C_j=b_{j}/a_{k}(k_{j})$ and $\bar{C}_j=\bar{b}_{j}/\bar{a}_{k}(\bar{k}_{j})$ we consider their numerators and denominators separately. For the denominators we employ trace formulae which can be used to show that the data  $a_{k}(k_j),\bar{a}_{k}(\bar{k}_j)$ depend on the eigenvalues $k_j, \bar{k}_j.$ For the numerators, it turns out that we can find symmetries relating $b_j,\bar{b}_j$. 

\subsection{Trace formulae}

\subsubsection{Space-time shifted reduction}
The derivation of the trace formulae follows an identical set of steps as shown in \cite{AM_Studies}. As such, the details are omitted. We 
assume that $a(k)$ and $\bar{a}(k)$ have simple zeros $\{k_{j}: \Im k_{j}>0\}_{j=1}^{J}$ and $\{\bar{k}_{j}: \Im \bar{k}_{j}<0\}_{j=1}^{J}$ respectively. 
The following trace formulae can be obtained:
\begin{eqnarray}
\label{Trace-a-cont}
\log a(k)&\underset{\Im k>0}{=}&\sum_{j=1}^{J} \log\left(\frac{k-k_{j}}{k-\bar{k}_{j}}\right) + \frac{1}{2\pi i}\int_{\mathbb{R}}
\frac{\log [1 +  b(t,\xi) \bar{b}(t,\xi) ]}{\xi-k}d\xi,\\
\label{Trace-a-bar-cont}
\log \bar{a}(k)&\underset{\Im k<0}{=}&\sum_{j=1}^{J} \log\left(\frac{k-\bar{k}_{j}}{k-{k}_{j}}\right) - \frac{1}{2\pi i}\int_{\mathbb{R}}
\frac{\log [1 +  b(t,\xi) \bar{b}(t,\xi) ]}{\xi-k}d\xi.
\end{eqnarray}
We shall restrict the discussion to reflectionless potentials for which $b= \bar{b}=0.$ In this case, the second terms in \eqref{Trace-a-cont} and \eqref{Trace-a-bar-cont} vanish. If we assume we have only a single eigenvalue $k=k_1, \bar{k}_1 \in \mathbb{C}$, we find
\begin{equation}
\label{Trace-a-der-1-cont}
a_{k}(k_1)= \frac{1}{k_{1} - \bar{k}_{1}}
 \;, \;\;\;\;\; 
 \bar{a}_{k}(\bar{k}_1)=   \frac{1}{\bar{k}_1 - k_1} \;.
\end{equation}
\subsubsection{Space shifted reduction and conjugate space-time shifted reduction}
In the space shifted case, as well as the conjugate space-time shifted case associated with the mKdV system, we found that the zeros of $a$ and $\bar{a}$ appear in pairs.
Denote by
$\{ k_{j}, -k_{j}^{*}:\Im k_{j}>0\}_{j=1}^{J}$ and $\{\bar{k}_{j}, -\bar{k}_{j}^{*}: \Im \bar k_{j}<0\}_{j=1}^{J}$, the (simple) zeros of
$a(k)$ and $\bar{a}(k)$ respectively. The general trace formulae are:
\begin{eqnarray}
\log a(k)&\underset{\Im k>0}{=}&\sum_{j=1}^{J} \log\left[\frac{(k-{k}_{j})(k+{k}_{j}^{*})}{(k-\bar k_{j})(k+\bar k_{j}^{*})}\right] + \frac{1}{2\pi i}\int_{\mathbb{R}}
\frac{\log [1 +  b(t,\xi) \bar{b}(t,\xi) ]}{\xi-k}d\xi,\\
\log \bar{a}(k)&\underset{\Im k<0}{=}&\sum_{j=1}^{J} \log\left[\frac{(k-\bar{k}_{j})(k+\bar{k}_{j}^{*})}{(k-k_{j})(k+k_{j}^{*})}\right] - \frac{1}{2\pi i}\int_{\mathbb{R}}
\frac{\log [1 +  b(t,\xi) \bar{b}(t,\xi) ]}{\xi-k}d\xi.
\end{eqnarray}
Again, we consider reflectionless potentials and assume a single eigenvalue pair $k=k_1,-k_1^{*}, \bar{k}_1, -\bar{k}_1^{*}$, in which case we find
\begin{eqnarray}
\label{Trace-a-der-PT-eig1-cont}
&a_{k}(k_1) =  \frac{k_{1}+k_{1}^{*}}{(k_{1}-\bar{k}_1)(k_1+\bar{k}_1^{*})}
\;,\;\;\;a_{k}(-k^*_1) =  \frac{k_{1}^{*}+k_{1}}{(k_1^{*} + \bar{k}_1)(-k_1^{*}+\bar{k}_1^{*})}
\;,\\
\label{Trace-a-ba-der-PT-eig1-cont}
&\bar{a}_{k}(\bar{k}_1) =  
\frac{\bar{k}_{1}+\bar{k}_{1}^{*}}{(\bar{k}_1 - k_1)(\bar{k}_1 - k_1^{*})}
\;,\;\;\;\bar{a}_{k}(-\bar{k}^*_1) =  
\frac{\bar{k}_{1}^{*}+\bar{k}_{1}}{(\bar{k}_1^{*} + k_1)(-\bar{k}_1^{*} + k_1^{*})}
\;.
\end{eqnarray}
Later, for simplicity, we will consider the case when $k_{1},\bar{k}_{1}$ are on the imaginary axis. With this restriction, the calculations become identical to the space-time shift case, and the formulae for $a_{k}$ and $\bar{a}_{k}$ given in (\ref{Trace-a-der-1-cont}) hold.

\subsection{Computing symmetries of $b_j$ and $\bar{b}_j$}
In this section we obtain the symmetries that $b_j$ and $\bar{b}_j$ satisfy. 
This in turn will be later used to determine the dependence of the norming constants $C_j$ and $\bar{C}_j$ on these eigenvalues. 

\subsubsection{Space-time shifted reduction}
At a discrete eigenvalue $k_{j}$, we have the relationship
\begin{equation}
\label{phi-linear-dependence-1-disc-eig-Jost-eig-2-cont}
M (x,t,k_j) = b_j(t) e^{2ik_{j}x} N (x,t,k_{j}) \;.
\end{equation}
Using the symmetry relation \eqref{nobar-jost-sym-sts}, together with the second component of Eq.~(\ref{phi-linear-dependence-1-disc-eig-Jost-eig-2-cont}), results in
\begin{equation}
\label{first-sym-summary-2c-N-and-M-prime-new-cond-comp-1b1-b-cont}
 N_{1} (x,t,k_j)  = -\sigma b_j(t_{0}-t) e^{2ik_{j}(x_{0}-x)} N_{2}(x_{0}-x,t_{0}-t,k_{j}) \;.
 \end{equation}
 Next, applying the symmetry relation
 (\ref{nobar-jost-sym-sts}) to the first component of
(\ref{phi-linear-dependence-1-disc-eig-Jost-eig-2-cont}) gives rise to
 \begin{equation}
\label{first-sym-summary-2c-N-and-M-prime-new-cond-comp-2b-b-cont}
  N_{2} (x_{0}-x,t_{0}-t,k_{j})     =  b_j(t) e^{2ik_{j}x} N_{1} (x,t,k_{j}) \;.
\end{equation}
Substituting \eqref{first-sym-summary-2c-N-and-M-prime-new-cond-comp-2b-b-cont} into \eqref{first-sym-summary-2c-N-and-M-prime-new-cond-comp-1b1-b-cont} one 
obtains the symmetry 
\begin{equation}
\label{b-sym-at-eig-val-RST-cont}
 b_j(t_{0}-t) b_j(t)
 =
 -\sigma  e^{-2ik_{j}x_{0}} \;
\;\;\;\;\; \Longrightarrow \;\;\;\;\;
 b^2_j(t_{0}/2) = -\sigma  e^{-2ik_{j}x_{0}}
  \;,
 \end{equation} 
where we used the time evolution of $b(t)$ given by (\ref{time-evolv-a-and-b-explicit_cont}). For a one soliton solution we find
 \begin{equation} 
 \label{b1-one-sol-cont}
b_1(t_{0}/2) =  \gamma_{1}e^{i (1+\sigma )\pi /4}e^{-ik_{1}x_{0}}\;, \;\;\; \gamma_{1} = \pm 1\;.
 \end{equation} 
To determine the value of $\bar{b}_j(t)$, one follows similar steps, which lead to
\begin{equation}
\label{b-b-bar-RST-2-cont} 
 \bar{b}_j (t_{0}-t) \bar{b}_j (t) = - \sigma   e^{2i\bar k_{j}x_{0}} \;\;\;\; \Longrightarrow \;\;\;\;  
 \bar{b}^2_j (t_{0}/2) = - \sigma e^{2i\bar{k}_{j}x_{0}} \;.
\end{equation}
In obtaining the expression for $\bar{b}_j (t_{0}/2)$ we made use of the evolution of the scattering data $\bar{b}_j (t)$ given 
by (\ref{time-evolv-abar-and-bbar-explicit_cont}). For the one soliton case we find
 \begin{equation} 
 \label{b1-bar-one-sol-cont}
\bar{b}_1(t_{0}/2) =   \bar\gamma_{1}e^{i (1+\sigma )\pi /4}e^{ik_{1}x_{0}} \;\;,\;\;\; \bar\gamma_{1}= \pm 1 \;.
 \end{equation} 
With the help of (\ref{Trace-a-der-1-cont}) and \eqref{C-ell-evolve_cont-shifted} we are now ready to compute the norming constants $C_1$ and $\bar{C}_1$ for all time. 
We thus have 
\begin{eqnarray}
\label{norming-const-C1-C1-bar-final-cont}
C_{1}(t)&=&\gamma_{1}e^{i (1+\sigma )\pi /4}(k_{1}-\bar{k}_{1})e^{-ik_{1}x_{0}}e^{2ik_{1}^{2}t_{0}}e^{-4ik_{1}^{2}t}
 \;,\\
 \label{norming-const-C1-C1-bar-final_2-cont}
\bar C_{1}(t)&=&\bar\gamma_{1}e^{i (1+\sigma )\pi /4}(\bar k_{1}-{k}_{1})e^{i\bar k_{1}x_{0}}e^{-2i\bar k_{1}^{2}t_{0}}e^{4i\bar k_{1}^{2}t}\;.
\end{eqnarray}

\subsubsection{Space shifted reduction}
Our starting point is 
again Eq.~(\ref{phi-linear-dependence-1-disc-eig-Jost-eig-2-cont}). Substituting the symmetry condition 
(\ref{nobar-jost-sym-ss}) into the second component one finds
\begin{equation}
\label{first-sym-summary-2c-N-and-M-prime-new-cond-comp-1b1-b-PT-cont}
 N_{1}(x,k_j)  = -\sigma  b_j^{*} e^{2ik_{j}(x_{0}-x)} N_{2}^{*}(x_{0}-x,-k_j^{*}) \;.
 \end{equation}
 Next, we use the symmetry (\ref{nobar-jost-sym-ss})
to rewrite the first component of Eq.~(\ref{phi-linear-dependence-1-disc-eig-Jost-eig-2-cont}) in the form
 \begin{equation}
\label{first-sym-summary-2c-N-and-M-prime-new-cond-comp-2b-b-PT-cont}
  N_{2}^{*}(x_{0}-x,-k_j^{*})     =  b_j e^{2ik_{j}x} N_{1} (x,k_j)  \;.
\end{equation}
Substituting Eq.~(\ref{first-sym-summary-2c-N-and-M-prime-new-cond-comp-2b-b-PT-cont}) back 
into (\ref{first-sym-summary-2c-N-and-M-prime-new-cond-comp-1b1-b-PT-cont}) and taking the one-soliton case gives
\begin{equation}
|b_1|^2   = -\sigma e^{-2ik_{1}x_{0}}.
 \end{equation}
This is only possible if $\sigma=-1$ and $k_{1}$ is on the imaginary axis, i.e. $k_{1}\equiv i\eta_{1}$ where $\eta_{1}\in\mathbb{R}^{+}$. In this case we have
\begin{equation}
    \label{b1-t-PT-1-cont}
    b_{1}=e^{i\theta_{1}}e^{\eta_{1}x_{0}},\;\;\;\theta_{1}\in\mathbb{R}.
\end{equation}
A similar expression can be obtained for $\bar{b}_j.$ Indeed, following an analogous procedure one can find
\begin{equation}
\label{b1-t-PT-bar-cont}
|\bar{b}_1 |^2   = -\sigma e^{2ik_{1}x_{0}}
   \;.
 \end{equation}
 Again, we take $\sigma=-1$ and $\bar{k}_{1}\equiv-i\bar\eta_{1}$ with $\bar\eta_{1}\in\mathbb{R}^{+}$, leading to
\begin{equation}
\label{b1-t-PT-bar-2-cont}
\bar{b}_1   = e^{i\bar{\theta}_1}e^{\bar\eta_{1}x_{0}}  \;,\;\;\; \bar{\theta}_{1}\in\mathbb{R}\;.
 \end{equation}
To write down a closed form expression for the norming constants $C_1$ and $\bar{C}_1$ we use the above formulae, the definition and time-evolution of the norming 
constants, and Eq.~(\ref{Trace-a-der-1-cont}):
\begin{eqnarray}
\label{norming-const-C1-C1-bar-final-PT-cont}
C_1(t) &=&  ie^{i \theta_1}(\eta_1 + \bar{\eta}_1)e^{\eta_{1}x_{0}} e^{4i\eta_{1}^{2}t}
 \;,\\
 \label{norming-const-C1-C1-bar-final-PT-2-cont}
\bar{C}_1(t) &=&  -ie^{i \bar{\theta}_1}({\eta}_1 +\bar\eta_1)e^{\bar\eta_{1}x_{0}}e^{-4i\bar \eta_{1}^{2}t} \;.
\end{eqnarray}

\subsubsection{Conjugate space-time shifted reduction}
Recall that this symmetry is associated with the conjugate space-time shifted mKdV equation. We start by applying the symmetry \eqref{nobar-jost-sym-csts} to the second component of \eqref{phi-linear-dependence-1-disc-eig-Jost-eig-2-cont} gives
\begin{equation}
\label{csts-b-1}
    N_{1}(x,t,k_{j})=-\sigma b_{j}^{*}(t_{0}-t)e^{2ik_{j}(x_{0}-x)}N_{2}^{*}(x_{0}-x,t_{0}-t,-k_{j}^{*}) .
\end{equation}
 Now, applying \eqref{nobar-jost-sym-csts} to the first component of \eqref{phi-linear-dependence-1-disc-eig-Jost-eig-2-cont} we also have
\begin{equation}
    \label{csts-b-2}
    N_{2}^{*}(x_{0}-x,t_{0}-t,-k_{j}^{*})=b_{j}(t)e^{2ik_{j}x}N_{1}(x,t,k_{j}).
\end{equation}
Substituting \eqref{csts-b-2} into \eqref{csts-b-1} we find
\begin{equation}
   \label{csts-b-3}
    b_{j}(t)b_{j}^{*}(t_{0}-t)=-\sigma e^{-2ik_{j}x_{0}}.
\end{equation}
Unlike the previous reductions, here we apply the mKdV time evolution of $b$ given by\\ $b(t,k)=e^{8ik^{3}(t-t_{0}/2)}b(t_{0}/2,k)$, which for the one-soliton case leads to
\begin{equation}
    \label{csts-b-4}
    |b_{1}(t_{0}/2)|^{2}=-\sigma e^{-2ik_{1}x_{0}}.
\end{equation}
This is only possible if $\sigma=-1$ and $k_{1}$ is on the imaginary axis, i.e. $k_{1}\equiv i\eta_{1}$ with $\eta_{1}\in\mathbb{R}^{+}$. Under these assumptions, we have
\begin{equation}
    \label{csts-b-5}
b_{1}(t_{0}/2)=e^{i\theta_{1}}e^{\eta_{1}x_{0}},\;\;\;\;\theta_{1}\in\mathbb{R}.
\end{equation}
Following an analogous series of steps, one can obtain
\begin{equation}
    \label{csts-bbar-1}
    |\bar{b}_{1}(t_{0}/2)|^{2}=-\sigma e^{2i\bar{k}_{1}x_{0}}.
\end{equation}
Taking $\sigma=-1$ and $\bar{k}_{1}\equiv -i\bar\eta_{1}$ with $\eta_{1}\in\mathbb{R}^{+}$,
\begin{equation}
\label{csts-bbar-2}
\bar{b}_{1}(t_{0}/2)=e^{i\bar\theta_{1}}e^{\bar\eta_{1} x_{0}},\;\;\;\;\bar\theta_{1}\in\mathbb{R}.
\end{equation}
To write down a closed form expression for the norming constants $C_1$ and $\bar{C}_1$, we use the above formulae, the definition and time-evolution of the norming 
constants, and Eq.~(\ref{Trace-a-der-1-cont}):
\begin{eqnarray}
\label{csts-norming-nobar}
C_{1}(t)&=&ie^{i\theta_{1}}(\eta_{1}+\bar\eta_{1})e^{\eta_{1}x_{0}}e^{-4\eta_{1}^{3}t_{0}}e^{8\eta_{1}^{3}t},\\
\label{csts-norming-bar}
\bar{C}_{1}(t)&=&-ie^{i\bar\theta_{1}}(\eta_{1}+\bar\eta_{1})e^{\bar\eta_{1} x_{0}}e^{-4\bar\eta_{1}^{3}t_{0}}e^{8\bar\eta_{1}^{3}t}.
\end{eqnarray}

\section{One-soliton solutions -- Continuous case}

In this section we compute 1-soliton solutions for the time, space-time, and space shifted NLS equations, as well as the complex space-time shifted mKdV equation. Before considering the relevant nonlocal reductions, one can obtain the general 1-soliton for the coupled $(q,r)$ system by setting $J=\bar{J}=1$ and $\rho=\bar\rho=0$ in \eqref{N-Nbar-sys-1}-\eqref{N-Nbar-sys-2}, which are simply an algebraic system of equations under these assumptions. Once the relevant eigenfunctions $\bar N_{1}(x,t,\bar k_{1})$ and $N_{2}(x,t,k_{1})$ are known, they may be substituted into \eqref{r-recon} and \eqref{q-recon} to find:
\begin{eqnarray}
\label{r-sol}
    r(x,t)&=&\frac{-2i(k_{1}-\bar{k}_{1})^{2}{C}_{1}(t)e^{2i{k}_{1}x}}{(k_{1}-\bar{k}_{1})^{2}+C_{1}(t)\bar{C}_{1}(t)e^{2i(k_{1}-\bar{k}_{1})x}},\\
\label{q-sol}
    q(x,t)&=&\frac{2i(k_{1}-\bar{k}_{1})^{2}\bar{C}_{1}(t)e^{-2i\bar{k}_{1}x}}{(k_{1}-\bar{k}_{1})^{2}+C_{1}(t)\bar{C}_{1}(t)e^{2i(k_{1}-\bar{k}_{1})x}}.
\end{eqnarray}
Recall that these formulae apply to both the NLS and mKdV $(q,r)$ systems, and the difference will appear in the time evolution of the norming constants. To determine the 1-soliton solutions for each specific reduction, we apply the relevant symmetries in scattering space to the general form \eqref{q-sol}.

\subsection{Time shifted reduction}
In this case, we have symmetries connecting the ``upper" scattering data to the ``lower" scattering data. Particularly, from \eqref{eig-sym-cont} we have $\bar k_{1}=-k_{1}$ with $k_{1}\in\mathbb{C}$ and from \eqref{a-sym-ts} and \eqref{b-sym-ts} together with the definition of the norming constants \eqref{norming-cont} we obtain
\begin{equation}
    \label{norming-sym-ts-cont}
    \bar{C}_{1}(t)=-\sigma C_{1}(t_{0}-t)=-\sigma C_{1}(t_{0}/2)e^{-2ik_{1}^{2}t_{0}}e^{4ik_{1}^{2}t}.
\end{equation}
Substituting \eqref{norming-sym-ts-cont} into \eqref{q-sol} gives
\begin{equation}
\label{q-sol-ts}
q(x,t)=\frac{-2i\sigma C_{1}(t_{0}/2)e^{-2ik_{1}^{2}t_{0}}e^{4ik_{1}^{2}t}e^{2ik_{1}x}}{1-\sigma\frac{C_{1}(t_{0}/2)^{2}}{4k_{1}^{2}}e^{4ik_{1}x}}.
\end{equation}
This soliton solution has two arbitrary parameteres $k_{1}$ and $C_{1}(t_{0}/2)$ which in general may be complex. We note that for certain specific initial conditions, the denominator of \eqref{q-sol-ts} can vanish (for all time); and as such, these should be excluded from the set of admissible initial conditions.

\subsection{Space-time shifted reduction}
Recall that unlike the time-shifted case, here the ``upper" and ``lower" scattering data are not connected. Instead, $k_{1}$ and $\bar{k}_{1}$ are free parameters and the norming constants can be expressed explicitly in terms of these two parameters using the formulae given in \eqref{norming-const-C1-C1-bar-final-cont}-\eqref{norming-const-C1-C1-bar-final_2-cont}. Substituting these into \eqref{q-sol}, we obtain the 1-soliton solution
\begin{equation}
\label{q-sol-sts}
q(x,t)=\frac{-2ie^{i(1+\sigma)\pi/4}(k_{1}-\bar{k}_{1})e^{-2i\bar{k}_{1}x}e^{4i\bar{k}_{1}^{2}t}}{\bar\gamma_{1}e^{-i\bar{k}_{1}x_{0}}e^{2i\bar{k}_{1}^{2}t_{0}}+\sigma\gamma_{1}e^{-ik_{1}x_{0}}e^{2ik_{1}^{2}t_{0}}e^{2i(k_{1}-\bar{k}_{1})x}e^{-4i(k_{1}^{2}-\bar{k}_{1}^{2})t}},
\end{equation}
where we have used the fact that $e^{i(1+\sigma)\pi/2}=-\sigma$. The parameters $k_{1},\bar{k}_{1}$ are free complex constants defined in the upper/lower half plane respectively, and $\gamma_{1},\bar\gamma_{1}$ can be chosen arbitrarily as either $\pm1$. We remark that while no choice of parameters will cause the denominator to vanish for all time, this soliton can develop singularity in finite time.

\subsection{Space shifted reduction}
While the previous two reductions allow for 1-soliton solutions for either choice of $\sigma$ and complex eigenvalues, in the space shifted case we have the restrictions that $\sigma=-1$ and $k_{1}=i\eta_{1},\;\bar{k}_{1}=-i\bar\eta_{1}$ with $\eta_{1},\bar\eta_{1}\in\mathbb{R}^{+}$ (see Eq.\eqref{first-sym-summary-2c-N-and-M-prime-new-cond-comp-1b1-b-PT-cont}). With this, substituting \eqref{norming-const-C1-C1-bar-final-PT-cont}-\eqref{norming-const-C1-C1-bar-final-PT-2-cont} into \eqref{q-sol} gives
\begin{equation}
\label{q-sol-ss}
q(x,t)=\frac{2(\eta_{1}+\bar\eta_{1})e^{-2\bar\eta_{1}x}e^{-4i\bar\eta_{1}^{2}t}}{e^{-i\bar\theta_{1}}e^{-\bar\eta_{1}x_{0}}-e^{i\theta_{1}}e^{\eta_{1}x_{0}}e^{-2(\eta_{1}+\bar\eta_{1})x}e^{4i(\eta_{1}^{2}-\bar\eta_{1}^{2})t}} \;.
\end{equation}
This soliton has four arbitrary real parameters $\eta_{1},\bar{\eta}_{1},\theta_{1},\bar\theta_{1}$, and like the space-time shifted case, can develop singularity in finite time. By renaming the arbitrary constants $\bar\eta_{1}\leftrightarrow-\eta_{1}$, $\bar\theta_{1}\leftrightarrow-\theta_{1}$, one can see that the present result agrees exactly with the one previously published in \cite{AM_shifted}.

\subsection{Conjugate space-time shifted reduction (mKdV)}
While the previous three cases correspond to soliton solutions of shifted nonlocal NLS type equations, we now construct the 1-soliton for the conjugate space-time shifted mKdV equation \eqref{space-time-shifted-complex-mkdv}. Substituting \eqref{csts-norming-nobar}-\eqref{csts-norming-bar} into \eqref{q-sol}, we have
\begin{equation}
    \label{q-sol-csts}
    q(x,t)=\frac{2(\eta_{1}+\bar\eta_{1})e^{-2\bar\eta_{1}x}e^{8\bar\eta_{1}^{3}t}}{e^{-i\bar\theta_{1}}e^{-\bar\eta_{1}x_{0}}e^{4\bar\eta_{1}^{3}t_{0}}-e^{i\theta_{1}}e^{\eta_{1}x_{0}}e^{-4\eta_{1}^{3}t_{0}}e^{-2(\eta_{1}+\bar\eta_{1})x}e^{8(\eta_{1}^{3}+\bar\eta_{1}^{3})t}}.
\end{equation}
This soliton has four arbitrary real parameters $\eta_{1},\bar\eta_{1},\theta_{1},\bar\theta_{1}$. 
\section{Direct scattering -- Discrete case}
\subsection{Linear Pair}
We start by considering the Ablowiz-Ladik (AL) scattering problem 
\begin{eqnarray}
\label{disc-scat3}
v_{n+1}&=&\mathcal{N}v_{n},\;\;\mathcal{N}=\begin{pmatrix}
z& Q_{n}(t)\\
R_{n}(t)& z^{-1}
\end{pmatrix},\\
\label{E:time}
\partial_{t}v_{n}&=&
\mathcal{T}v_{n},\;\;\mathcal{T}=\begin{pmatrix}
 i Q_{n}(t)R_{n-1}(t)-\frac{i}{2}(z-z^{-1})^{2}& -i(zQ_{n}(t)-z^{-1}Q_{n-1}(t))\\
 i(z^{-1}R_{n}(t)-zR_{n-1}(t))& - i R_{n}(t)Q_{n-1}(t)+\frac{i}{2}(z-z^{-1})^{2}
\end{pmatrix}.
\end{eqnarray}
where $v_{n}(t)\equiv (v_{n}^{(1)}(t), v_{n}^{(2)}(t))^{T}$ is a complex valued function of 
$t\in\mathbb{R}$
 and we take $n\in\mathbb{Z}$. Furthermore, $Q_n(t)$ and $R_n(t)$ are complex valued potentials that are assumed to rapidly decay to zero as $n\rightarrow\pm\infty.$  Here, $z$ is a spectral parameter taken to be (in general) complex. The discrete compatibility condition $\partial_{t}v_{n+1}=(\partial_{t}v_m)_{m=n+1}$ yields the system of equations (\ref{discrete_qr1}) and (\ref{discrete_qr2}).
Due to the spatially nonlocal nature of the symmetry reductions (\ref{RQ2}) and (\ref{RQ3}), it proves crucial for obtaining the symmetries in scattering space to supplement the AL system (\ref{disc-scat3}) with a ``backward" scattering problem (obtained by 
inverting Eq.~(\ref{disc-scat3})) defined by
\begin{equation}
\label{inv_scatt_prob}
w_{n-1}=
\tilde{\mathcal{N}}w_{n},\;\;\tilde{\mathcal{N}}=\begin{pmatrix}
z^{-1}& -Q_{n}(t)\\
- R_{n}(t)& z
\end{pmatrix}.
\end{equation}
It was shown in \cite{AM_discrete} that any solution $v_{n}(z,t)$ of (\ref{disc-scat3}) can be related  to a solution of (\ref{inv_scatt_prob}) via the transformation
\begin{equation}
\label{vnwn-relate}
w_{n}(z,t)=f_{n}(t)v_{n+1}(z,t), \;\;\; f_{n}(t) \equiv \prod_{k=-\infty}^{n}\frac{1}{1 - Q_{k}(t) R_{k}(t)} \;.
\end{equation}
\subsection{Eigenfunctions and scattering data}
In this section, we provide the main ingredients necessary to solve the AL scattering problem for generic potentials. 
Since the discrete potentials $Q_n, R_n$ vanish rapidly as $n\rightarrow\pm\infty$, we may define the following eigenfunctions by their boundary conditions \cite{APT_Book}:
\begin{eqnarray}
\label{BC1}
&\phi_{n}(z,t)\sim z^{n}\mathbf{\hat{e}_{1}}, \ \ \
&\bar{\phi}_{n}(z,t)\sim z^{-n}\mathbf{\hat{e}_{2}},\;\;\text{as}\;\;n\rightarrow-\infty,\\
\label{BC2}
&\psi_{n}(z,t)\sim z^{-n}\mathbf{\hat{e}_{2}}, \ \ \
&\bar{\psi}_{n}(z,t)\sim z^{n}\mathbf{\hat{e}_{1}},\;\,\;\;\text{as}\;\;n\rightarrow+\infty.
\end{eqnarray}
Clearly, the eigenfunction pairs $\{\phi_{n}, \bar{\phi}_{n}\}$ and $\{\psi_{n},  \bar{\psi}_{n}\}$ are two linearly independent sets of solutions, so we can write
\begin{eqnarray}
\label{phi-linear-dependence-1-disc}
\phi_{n}(z,t)&=&a(z,t)\bar{\psi}_{n}(z,t)+b(z,t)\psi_{n}(z,t) ,\\
\label{phi-bar-linear-dependence-1-disc}
\bar{\phi}_{n}(z,t)&=&\bar{a}(z,t)\psi_{n}(z,t)+\bar{b}(z,t)\bar{\psi}_{n}(z,t) ,
\end{eqnarray}
where $a(z,t), \bar{a}(z,t), b(z,t), \bar{b}(z,t)$ are the scattering data given by the relations
\begin{eqnarray}
\label{W-a}
a(z,t) &=& c_n(t) W(\phi_n (z,t),\; \psi_n (z,t)) ,\\
\label{W-a-bar}
\bar{a}(z,t) &=& c_n(t) W(\bar{\psi}_n (z,t),\; \bar{\phi}_n (z,t)) ,\\
\label{W-b}
b(z,t) &=& c_n(t) W(\bar{\psi}_n (z,t),\; \phi_n (z,t)) , \\
\label{W-b-bar}
\bar{b}(z,t) &=& c_n(t) W(\bar{\phi}_n (z,t),\; \psi_n (z,t)) .
\end{eqnarray}
Here, $W$ is the Wronskian defined by
$W(v_n,w_n)= v^{(1)}_{n} w^{(2)}_{n}  -  v^{(2)}_{n} w^{(1)}_{n}$
where $v_n = (v^{(1)}_{n},\; v^{(2)}_{n})^T$ and $w_n = (w^{(1)}_{n},\; w^{(2)}_{n})^T$ and 
\begin{equation}
\label{c-n}
c_n(t)  =   \prod_{k=n}^{+\infty} \left( 1-Q_k(t) R_k (t)\right).
\end{equation}
The above scattering data also satisfies the condition 
\begin{equation}
\label{W-phi-psi-relation4}
a(z,t) \bar{a}(z,t) - b(z,t) \bar{b}(z,t)  = c_{-\infty} .
\end{equation}
As we shall see later, the scattering data $a, \bar{a}$ and the product $ b\bar{b}$ turn out to be time-independent 
(see \cite{APT_Book} for further details). This fact implies that $c_{-\infty}$ is time-independent as well, thus making it a constant of motion.
In the following analysis, it is convenient to consider functions with constant
boundary conditions. We define the bounded eigenfunctions as follows:
\begin{equation}
M_{n}(z,t)=z^{-n}\phi_{n}(z,t), \ \ \bar{M}_{n}(z,t)=z^{n}\bar{\phi}_{n}(z,t), \ \ 
N_{n}(z,t)=z^{n}\psi_{n}(z,t), \ \ \bar{N}_{n}(z,t)=z^{-n}\bar{\psi}_{n}(z,t).
\end{equation}
Notably, it can be shown that $M_{n}$ and $N_{n}$ are analytic inside the unit circle $|z|=1$, while $\bar M_{n}$ and $\bar N_{n}$ are analytic outside the unit circle (see for example \cite{APT_Book}). Additionally, it proves convenient to modify these eigenfunctions and introduce instead a new set of eigenfunctions
\begin{equation}
    \label{mod_efns}
M_{n}'(z,t)=\mathcal{A}M_{n}(z,t),\;\;N_{n}'(z,t)=\mathcal{A}N_{n}(z,t),\;\;\bar M_{n}'(z,t)=\mathcal{A}\bar M_{n}(z,t),\;\;\bar N_{n}'(z,t)=\mathcal{A}\bar N_{n}(z,t),
\end{equation}
where $\mathcal{A}$ denotes the matrix
\begin{equation}
\label{A-matrix}
\mathcal{A}
\equiv 
 \begin{pmatrix}
1& 0\\  
0& c_{n}(t)
\end{pmatrix}
.
\end{equation}
We will use the set of eigenfunctions \eqref{mod_efns} in the formulation of the inverse problem later on. Similarly to the continuous case, one can write \eqref{phi-linear-dependence-1-disc} and \eqref{phi-bar-linear-dependence-1-disc} as
\begin{eqnarray}
    \label{rh1d}
    \mu_{n}'(z,t)&=&\bar{N}_{n}'(z,t)+\rho(z,t)z^{-2n}N_{n}'(z,t),\\
    \label{rh2d}
    \bar\mu_{n}'(z,t)&=&N_{n}'(z,t)+\bar\rho(z,t)z^{2n}\bar{N}_{n}'(z,t),
\end{eqnarray}
where $\mu_{n}'=M_{n}'/a$, $\bar\mu_{n}'=\bar{M}_{n}'/\bar{a}$, $\rho=b/a$, and $\bar\rho=\bar{b}/\bar{a}$. Eventually, \eqref{rh1d} and \eqref{rh2d} will be viewed as a Riemann-Hilbert problem on the unit circle in the complex $z$-plane.
\subsection{Relation between the scattering data for the spatially shifted and unshifted AL systems}
We now derive the connection between the shifted and unshifted scattering data alluded to in section 2.2. Denote by $\Phi_m,\bar\Phi_m,\Psi_m,\bar\Psi_m$ the eigenfunctions associated with the unshifted nonlocal scattering problem (\ref{disc-scat3}) with $R_m(t)=\sigma Q^*_{-m}(t)$, given by
\begin{eqnarray}
\label{eigenphi-un-dis}
&\Phi_{m}(z)\sim z^{m}\mathbf{\hat{e}_{1}},\;\;\;\;&\bar{\Phi}_{m}(z)\sim z^{-m}\mathbf{\hat{e}_{2}},\;\;\;\;\;\;\;\text{as}\;m\rightarrow-\infty,\\
\label{eigenpsi-un-dis}
&\Psi_{m}(z)\sim z^{-m}\mathbf{\hat{e}_{2}},\;\;\;\;&\bar\Psi_{m}(z)\sim z^{m}\mathbf{\hat{e}_{1}},\;\;\;\;\;\;\;\;\;\text{as}\;m\rightarrow+\infty.
\end{eqnarray}
Letting $m=n-n_{0}/2$ and comparing (\ref{eigenphi-un-dis}) - (\ref{eigenpsi-un-dis}) with 
\eqref{BC1}-\eqref{BC2} gives the relationships
\begin{eqnarray}
\label{unshift1-dis}
    &\phi_{n}(z)=z^{n_{0}/2}\Phi_{n-n_{0}/2}(z),\;\;\;\;&\bar\phi_{n}(z)=z^{-n_{0}/2}\bar\Phi_{n-n_{0}/2}(z),\\
    \label{unshift2-dis}
     &\psi_{n}(z)=z^{-n_{0}/2}\Psi_{n-n_{0}/2}(z),\;\;\;\;&\bar\psi_{n}(z)=z^{n_{0}/2}\bar\Psi_{n-n_{0}/2}(z).
\end{eqnarray}
Furthermore, let $A,B,\bar{A},\bar{B}$ denote the scattering data associated with the unshifted problem. The Wronskian relations \eqref{W-a}-\eqref{W-b-bar} together with \eqref{unshift1-dis}-\eqref{unshift2-dis} lead to the conclusion
\begin{eqnarray}
\label{unshifta-dis}
    &a(z)=A(z),\;\;\;\;&\bar{a}(z)=\bar{A}(z),\\
    \label{unshiftb-dis}
    &b(z)=z^{n_{0}}B(z),\;\;\;\;&\bar{b}(z)=z^{-n_{0}}\bar B(z).
\end{eqnarray}
Note the similarity between the property of the semi-discrete Fourier transform given in \eqref{disc-ft-connection} and the relations in \eqref{unshiftb-dis}.
\section{Symmetries in scattering space -- Discrete case}
In this section, we apply the various shifted nonlocal symmetry reductions \eqref{RQ1}-\eqref{RQ3} and derive the corresponding symmetry relationships between the eigenfunctions and scattering data in each case. Throughout this section, we will make frequent use of the following definitions for ease of notation:
\begin{eqnarray}
&\Lambda_{\pm}\equiv\begin{pmatrix}0&\pm\sigma\\1&0\end{pmatrix},\;\;\;\Lambda_{\pm}^{-1}\equiv\begin{pmatrix}0&1\\\pm\sigma&0\end{pmatrix},\\
    &\Gamma_{n}(t)\equiv\begin{pmatrix}c_{n}(t)^{-1}&0\\0&c_{n}(t)\end{pmatrix},\;\;\;\tilde\Gamma\equiv\begin{pmatrix}c_{-\infty}^{-1}&0\\0&1\end{pmatrix}.
\end{eqnarray}

\subsection{Time-shifted symmetry reduction}

\subsubsection{Symmetries between the eigenfunctions}
Let $v_n (z,t)\equiv (v^{(1)}_n (z,t),\; v^{(2)}_n (z,t))^{T}$ be any solution to system (\ref{disc-scat3}) with $R_n(t)=\sigma Q_n(t_{0}-t)$:
\begin{eqnarray}
\label{disc-v1-eqn-minus}
v^{(1)}_{n+1}(z,t)&=&zv^{(1)}_{n}(z,t)   +   Q_n(t)v^{(2)}_{n}(z,t),\\
\label{disc-v2-eqn-minus}
v^{(2)}_{n+1}(z,t)&=&  \sigma Q_n(t_{0}-t)v^{(1)}_{n}(z,t)  +  z^{-1}v^{(2)}_{n}(z,t).\nonumber
\end{eqnarray}
Make the change of variables $t\rightarrow t_{0}-t,\; z\rightarrow 1/z$ in the above equations to find
\begin{eqnarray}
\label{disc-v2-eqn-c-minus}
v^{(2)}_{n+1}(1/z,t_{0}-t) &=& z v^{(2)}_n(1/z,t_{0}-t) + Q_{n}(t)[\sigma v^{(1)}_n(1/z,t_{0}-t)],\\
\label{disc-v1-eqn-c-minus}
\sigma v^{(1)}_{n+1}(1/z,t_{0}-t) &=& \sigma Q_n(t_{0}-t) v^{(2)}_n(1/z,t_{0}-t) + z^{-1}[\sigma v^{(1)}_n(1/z,t_{0}-t)].\nonumber
\end{eqnarray}
Therefore, we have the following 
symmetry relation:
\begin{equation*}
\label{first-sym-summary-minus2}
{\rm If}\;  v_{n}(z,t)\;
{\rm solves}\;(\ref{disc-scat3})\; {\rm with}\; R_n(t)=\sigma Q_{n}(t_{0}-t) \; {\rm so~  does}\;
 \Lambda_{+}v_{n}(1/z,t_{0}-t).
\end{equation*}
Now let $\phi_{n}(z,t)$ be the solution to system (\ref{disc-scat3}) 
with $R_n(t)= \sigma Q_{n}(t_{0}-t)$ that obeys the boundary condition given on the left part of Eq.~(\ref{BC1}). From the above relation, we have that $\Lambda_{+}\phi_{n}(1/z,t_{0}-t)$ is also a solution.
Next, let $\bar{\phi}_n (z,t)$ be another solution to system (\ref{disc-scat3}) with $R_n(t)=\sigma Q_{n}(t_{0}-t)$ satisfying the boundary condition given 
on the right part of Eq.~(\ref{BC1}). Comparing boundary conditions, one can obtain the relationship:
\begin{equation}
\label{first-sym-summary-2c}
\bar{\phi}_n (z,t) =\Lambda_{+}{\phi}_n (1/z,t_{0}-t).
\end{equation}
Similarly, let
$\psi_n (z,t)$ 
be the solution to system (\ref{disc-scat3})  with $R_n(t)=\sigma Q_{n}(t_{0}-t)$ that satisfies the boundary condition (\ref{BC2}).
If $\bar{\psi}_n (z,t)$ is a solution to system (\ref{disc-scat3}), with $R_n(t)=\sigma Q_{n}(t_{0}-t),$ satisfying the boundary 
condition (\ref{BC2}) we have the result:
\begin{equation}
\label{first-sym-summary-2c-bar}
\bar{\psi}_{n} (z,t) =\Lambda_{+}^{-1}{\psi}_{n} (1/z,t_{0}-t).
\end{equation}

\subsubsection{Symmetries between the modified eigenfunctions}
These symmetry relations induce important symmetries between the eigenfunctions $\bar{M}_n(z,t)$ and $M_n(z,t)$ and between $\bar{N}_n(z,t)$ and $N_n(z,t)$ which read
\begin{eqnarray}
\label{first-sym-summary-2c-N-and-M2a}
\bar{M}_{n} (z,t)&=&
\Lambda_{+}M_{n} (1/z,t_{0}-t)  
\;,\\
\label{first-sym-summary-2c-N-and-M-bar20}
\bar{N}_{n} (z,t)&=&
\Lambda_{+}^{-1} {{N}_{n}} (1/z,t_{0}-t) 
\;.
\end{eqnarray}
Using the symmetry relation between the ``unprimed" eigenfunctions 
established in (\ref{first-sym-summary-2c-N-and-M2a}) and (\ref{first-sym-summary-2c-N-and-M-bar20}), we find the following symmetry condition for the ``primed" eigenfuncitons,
\begin{eqnarray}
\label{first-sym-summary-2c-N-and-M-prime-new-cond-comp-1}
 {\bar{M}'_{n}} (z,t)   &=&  \Gamma_{n}(t)\Lambda_{+}M'_{n} (1/z,t_{0}-t) \;,\\
\label{first-sym-summary-2c-N-and-M-prime-new-cond-M-bar-comp-1} 
{\bar{N}'_{n}} (z,t)   &=&  \Gamma_{n}(t)\Lambda_{+}^{-1}{{N}'_{n}} (1/z,t_{0}-t).
\end{eqnarray}

\subsubsection{Symmetries of the scattering data}
To establish the symmetry relation between the scattering data $a(z)$ and $\bar{a}(z)$ we start from (\ref{W-a}), (\ref{W-a-bar})
and the assumption that $R_n(t) = \sigma Q_n(t_{0}-t)$. Now we have
\begin{equation}
\label{W-a-1-bar}
\bar{a}(z,t) =
        c_n(t) [( {\psi}^{(2)}_{n} (1/z,t_{0}-t)  {\phi^{(1)}_{n}} (1/z,t_{0}-t)   - {\psi}^{(1)}_{n} (1/z,t_{0}-t)   {\phi^{(2)}_{n}} (1/z,t_{0}-t) ]
          .
\end{equation}
However, from (\ref{W-a}) we get
\begin{equation}
\label{W-a-3-bar}
a(1/z,t_{0}-t) =
c_n(t_{0}-t) [ {\psi}^{(2)}_{n} (1/z,t_{0}-t)  {\phi^{(1)}_{n}} (1/z,t_{0}-t)   -   {\psi}^{(1)}_{n} (1/z,t_{0}-t)   {\phi^{(2)}_{n}} (1/z,t_{0}-t) ] .
\end{equation}
Since $c_n(t) = c_n(t_{0}-t)$, we have
\begin{equation}
\label{discrete-scat-a-3}
\bar{a}(z,t) = a(1/z,t_{0}-t).
\end{equation}
As we shall see later, it turns out that the scattering data ${a}(z,t)$ and $\bar a(z,t)$ are time-independent, giving rise to the following 
symmetry between their zeros:
\begin{equation}
\label{eig-sym}
\text{If $a(z_j)=0$ and $\bar{z}_{j}=1/z_{j}$, then $\bar{a}(\bar{z}_{j})=0$.}
\end{equation}
To find the symmetry relation between the scattering data $b(z,t)$ and $\bar{b}(z,t)$
we start from Eq. (\ref{W-b-bar}) and use the symmetry relation (\ref{first-sym-summary-2c}) to obtain
\begin{equation}
\label{W-b-bar-1}
\bar{b}(z,t) =
            c_n(t) [\sigma {\phi}^{(2)}_n (1/z,t_{0}-t) \psi^{(2)}_n (z,t)  - {\phi}^{(1)}_n (1/z,t_{0}-t)\psi^{(1)}_n (z,t)].
\end{equation}
Next,  from (\ref{W-b}) we let $z\rightarrow 1/z, \; t\rightarrow t_{0}-t$ and make use of the symmetry relation (\ref{first-sym-summary-2c-bar}) to find
\begin{equation}
\label{W-b-1}
b(1/z,t_{0}-t) 
=c_n(t_{0}-t) [{\psi}^{(2)}_n (z,t) \phi^{(2)}_n (1/z,t_{0}-t)  - \sigma {\psi}^{(1)}_n (z,t)\phi^{(1)}_n (1/z,t_{0}-t)].
          \end{equation}
Comparing, we deduce the relationship
\begin{equation}
\label{discrete-scat-b-b-bar-RT}
\bar{b}(z,t) = \sigma b(1/z,t_{0}-t)\;.
\end{equation}
Note that $b$ and $\bar{b}$ are only defined for $z$ on the unit circle.
\subsection{Space-time shifted reduction}
\subsubsection{Symmetries between the eigenfunctions}
Assume that $v_n (z,t)\equiv (v^{(1)}_n (z,t),\; v^{(2)}_n (z,t))^{{T}}$  
is any solution to system (\ref{disc-scat3}) with $R_n(t)=  \sigma Q_{n_{0}-n}(t_{0}-t)$:
\begin{eqnarray}
\label{disc-v1-eqn-minus1}
v^{(1)}_{n+1}(z,t)&=&zv^{(1)}_{n}(z,t)   +   Q_n(t)v^{(2)}_{n}(z,t) ,\\
\label{disc-v2-eqn-minus1}
v^{(2)}_{n+1}(z,t)&=&  \sigma Q_{n_{0}-n}(z,t_{0}-t)v^{(1)}_{n}(z,t)  +  z^{-1}v^{(2)}_{n}(z,t) .\nonumber
\end{eqnarray}
Let $n\rightarrow n_{0}-n,\; t\rightarrow t_{0}-t$ in (\ref{disc-v1-eqn-minus1}) and (\ref{disc-v2-eqn-minus1}) to find
\begin{eqnarray}
\label{disc-v1-eqn-minus1a}
v^{(2)}_{n_{0}-(n-1)}(z,t_{0}-t) &=& z^{-1} v^{(2)}_{n_{0}-n}(z,t_{0}-t) +  \sigma Q_{n}(t) v^{(1)}_{n_{0}-n}(z,t_{0}-t) ,\\
\label{disc-v2-eqn-minus1a}
v^{(1)}_{n_{0}-(n-1)}(z,t_{0}-t) &=& Q_{n_{0}-n} (z,t_{0}-t) v^{(2)}_{n_{0}-n}(z,t_{0}-t) + z v^{(1)}_{n_{0}-n}(z,t_{0}-t) .\nonumber
\end{eqnarray}
 Thus, we can deduce the following symmetry relation:
\begin{equation*}
\label{first-sym-summary-minus-1}
{\rm If}\;\;  v_{n} (z,t)\;
{\rm solves}\;\; (\ref{disc-scat3})\; {\rm with}\; R_n(t)=\sigma Q_{n_{0}-n}(t_{0}-t) \;\; {\rm then}\;
 \Lambda_{-}v_{n_{0}-n}(z,t_{0}-t) \; 
{\rm solves}\; (\ref{inv_scatt_prob}).
\end{equation*}
Now, by definition $\phi_{n}(z,t)$ is a solution of \eqref{disc-scat3}, so we know $\Lambda_{-}\phi_{n}(z,t)$ is a solution of \eqref{inv_scatt_prob}.
We will compare this with the eigenfunction $\psi_{n}(z,t)$. Note that since $\gamma\psi_{n}(z,t)$ (where $\gamma$ is a constant to be determined) satisfies \eqref{disc-scat3}. Then, $\gamma f_n(t)\psi_{n+1}(z,t)$ solves \eqref{inv_scatt_prob}
 where $f_{n}(t)$ is as given in \eqref{vnwn-relate} with the appropriate symmetry reduction.
Comparing boundary conditions, we find that
\begin{equation}
\label{vn-wn-relation-sym-minus-new}
 {\phi^{(2)}_{n_{0}-n}} (z,t_{0}-t)  =\gamma f_{n}(t) \psi^{(1)}_{n+1} (z,t),\;\;\;\;\;\;\;
-\sigma {\phi^{(1)}_{n_{0}-n}} (z,t_{0}-t) = \gamma  f_{n}(t) \psi^{(2)}_{n+1} (z,t) ,
 \end{equation}
hold if we choose $\gamma =  -\sigma z^{n_{0}+1} c_{-\infty}$. With this, and using the fact that $c_{-\infty}f_{n}(t)=c_{n+1}(t)$, we have the following relationship between the eigenfunctions:
\begin{equation}
\label{first-sym-summary-between-eigfunc}
\psi_{n+1} (z,t) =\frac{1}{z^{n_{0}+1} c_{n+1}(t)}
 \Lambda_{-}{\phi_{n_{0}-n}} (z,t_{0}-t).
\end{equation}
Next, we derive the symmetry relations between the ``bar" eigenfunctions. Following similar steps as before, we compare the boundary condition satisfied by the eigenfunction $\Lambda_{-}^{-1}\bar\phi_{n_{0}-n}(z,t_{0}-t)$ with $\bar\gamma f_{n}(t)\bar\psi_{n+1}(z,t)$ (where $\bar\gamma$ is to be determined) after which we find that
\begin{equation}
\label{vn-wn-relation-sym-minus-new-bar-PT2}
  {\bar{\phi}^{(2)}_{n_{0}-n}} (z,t_{0}-t)  = \bar{\gamma}  f_{n}(t) \bar{\psi}^{(1)}_{n+1} (z,t),\;\;\;\;\;\;\;
-\sigma {\bar{\phi}^{(1)}_{n_{0}-n}} (z,t_{0}-t) = \bar{\gamma} f_{n}(t) \bar{\psi}^{(2)}_{n+1} (z,t),
 \end{equation}
 hold only when $\bar{\gamma} =  \frac{c_{-\infty}}{z^{n_{0}+1}}.$ Thus, the ``bar" eigenfunctions satisfy the following symmetry:
\begin{equation}
\label{first-sym-summary-between-eigfunc-bar}
\bar{\psi}_{n+1} (z,t)=
\frac{z^{n_{0}+1}}{ c_{n+1}(t)}\Lambda_{-}^{-1}\bar{\phi}_{n_{0}-n}(z,t_{0}-t).
\end{equation}
\subsubsection{Symmetries between the modified eigenfunctions}
The relations above in turn induce the following symmetries: 
\begin{eqnarray}
\label{first-sym-summary-2c-N-and-M-new}
N_{n+1} (z,t)  &=&
\frac{1}{c_{n+1}^{\sigma}(t)}
\Lambda_{-}{M_{n_{0}-n}} (z,t_{0}-t),\\
\label{first-sym-summary-2c-N-and-M-bar-new}
\bar{N}_{n+1} (z,t)&=&
\frac{1}{ c_{n+1}^{\sigma}(t)}
 \Lambda_{-}^{-1}{\bar{M}_{n_{0}-n}} (z,t_{0}-t) .
\end{eqnarray}
Applying these symmetry conditions to the ``primed" eigenfunctions and using the fact that $c_{n_{0}-n}(t_{0}-t)c_{n+1}(t)=c_{-\infty}$, one finds
\begin{eqnarray}
\label{sym-prime-RST-1}
 {N'_{n+1}} (z,t)  
  &=& \tilde\Gamma\Lambda_{-}{M'_{n_{0}-n}} (z,t_{0}-t) ,\\
{\bar{N}'_{n+1}} (z,t)     
&=& \tilde\Gamma\Lambda_{-}^{-1} {\bar{M}'_{n_{0}-n}} (z,t_{0}-t) .
\end{eqnarray}
\subsubsection{Symmetries of the scattering data}
We start from the definition given in (\ref{W-a}). Use the symmetry relation between the eigenfunctions obtained in (\ref{first-sym-summary-between-eigfunc}) to find
\begin{equation}
\label{W-a-30}
a(z,t) =
\frac{1}{z^{n_{0}+1} } [\phi^{(1)}_{n+1} (z,t) {\phi^{(1)}_{n_{0}-n}} (z,t_{0}-t)  
            +  \sigma \phi^{(2)}_{n+1} (z,t)  {\phi^{(2)}_{n_{0}-n}} (z,t_{0}-t) ]
          ,
\end{equation}
By letting $n\rightarrow n_{0}-n -1, t \rightarrow t_{0}-t$ in (\ref{W-a-30}) one finds that $a$ has the property
\begin{equation}
\label{discrete-scat-a}
a(z,t) ~ = ~  a(z,t_{0}-t) .
\end{equation}
A similar derivation for the scattering data $\bar{a}(z,t)$ is possible, leading to
\begin{equation}
\label{discrete-scat-a-33-RST}
\bar{a}(z,t) = \bar{a}(z,t_{0}-t) .
\end{equation}
Like in the continuous case, $a(z,t)$ and $\bar{a}(z,t)$ turn out to be time-independent, and as such Eqns.~(\ref{discrete-scat-a}) and (\ref{discrete-scat-a-33-RST})
are actually trivial. 
Next, we proceed with the derivation of the symmetry between $b(z,t)$ and $\bar{b}(z,t).$ We start from Eq. (\ref{W-b-bar}) and make use of the symmetry conditions between the eigenfunctions given in (\ref{first-sym-summary-2c-N-and-M-new}) and (\ref{first-sym-summary-2c-N-and-M-bar-new}) to obtain
\begin{equation}
\label{W-b-bar-100}
\bar{b}(z,t) =
  \frac{1 }{z^{n_{0}+1}}
  [\bar{\phi}^{(1)}_{n+1} (z)  {\phi^{(1)}_{n_{0}-n}} (z,t_{0}-t)  + \sigma \bar{\phi}^{(2)}_{n+1} (z,t) {\phi^{(2)}_{n_{0}-n}} (z,t_{0}-t) ] .
 \end{equation}
Next, from (\ref{W-b}) it follows that
\begin{equation}
\label{W-b-1000}
b(z,t)  =
 z^{n_{0}+1} 
 [\bar{\phi}^{(2)}_{n_{0}-n} (z,t_{0}-t) \phi^{(2)}_{n+1} (z,t)  +\sigma \bar{\phi}^{(1)}_{n_{0}-n} (z,t_{0}-t)\phi^{(1)}_{n+1} (z,t)]
  .
\end{equation}
Let $n\rightarrow n_{0}-n-1, t\rightarrow t_{0}-t$ in (\ref{W-b-1000}) to find
\begin{equation}
\label{discrete-scat-b-b-bar-RST}
\bar{b}(z,t) = \frac{\sigma}{z^{2(n_{0}+1)}}  b(z,t_{0}-t).
\end{equation}

\subsection{Space shifted reduction}

\subsubsection{Symmetries between the eigenfunctions}
Since this reduction is local in time, we shall omit the explicit time dependence from all dependent variables.
Assume that $v_n (z)\equiv (v^{(1)}_n (z),\; v^{(2)}_n (z))^{T}$ 
is any solution to system (\ref{disc-scat3}) with $R_n=\sigma Q^*_{n_{0}-n}.$ Taking the complex conjugate and making the transformation 
$n\rightarrow n_{0}-n,\;z\rightarrow z^*$ one obtains
\begin{eqnarray}
\label{disc-v2-eqn-c-minus1a}
v^{(2)*}_{n_{0}-(n-1)} (z^*) &=& z^{-1} v^{(2)*}_{n_{0}-n}(z^*) +\sigma Q_{n} v^{(1)*}_{n_{0}-n}(z^*) ,\\
\label{disc-v1-eqn-c-minus1a}
v^{(1)*}_{n_{0}-(n-1)}(z^*) &=& Q^*_{n_{0}-n} v^{(2)*}_{n_{0}-n}(z^*) + z v^{(1)*}_{n_{0}-n}(z^*) .\nonumber
\end{eqnarray}
 Therefore, we arrive at the following conclusion
\begin{equation*}
\label{first-sym-summary-minus}
{\rm If}\; v_{n} (z)\;
{\rm solves}\; (\ref{disc-scat3})\; {\rm with}\; R_n=\sigma Q^*_{n_{0}-n} \; {\rm then}\;
\Lambda_{-}v^{*}_{n_{0}-n}(z^*) \;
{\rm solves}\;\; (\ref{inv_scatt_prob}).
\end{equation*}
In a similar fashion to the space-time shifted case, it can be found that
\begin{equation}
\label{vn-wn-relation-sym-minus-1}
 {\phi^{(2)*}_{n_{0}-n}} (z^*) ~ = \gamma f_{n} \psi^{(1)}_{n+1} (z),\;\;\;\;\;
  {\phi^{(1)*}_{n_{0}-n}} (z^*) ~ = -\sigma \gamma f_{n} \psi^{(2)}_{n+1} (z),
 \end{equation}
holds if we require $\gamma =  -\sigma  c_{-\infty} z^{n_{0}+1}.$
Thus, using the fact that $c_{-\infty}f_{n}=c_{n+1}$, we have the following relationship between the eigenfunctions:
\begin{equation}
\label{first-sym-summary-between-eigfunc-PT}
\psi_{n+1} (z) =
\frac{1}{z^{n_{0}+1} c_{n+1}}
\Lambda_{-}{\phi^{*}_{n_{0}-n}} (z^*)   .
\end{equation}
To establish the symmetry relations between the ``bar" eigenfunctions, we find that
\begin{equation}
\label{vn-wn-relation-sym-minus-new-bar}
  {\bar{\phi}^{(2)*}_{n_{0}-n}} (z^*) = \bar{\gamma} f_{n} \bar{\psi}^{(1)}_{n+1} (z),\;\;\;\;\;\;\;
-\sigma {\bar{\phi}^{(1)*}_{n_{0}-n}} (z^*)  = \bar{\gamma} f_{n} \bar{\psi}^{(2)}_{n+1} (z) ,
 \end{equation}
holds if we require $\bar{\gamma} =  \frac{c_{-\infty}}{z^{n_{0}+1}}$.
Thus we have the following relationship:
\begin{equation}
\label{first-sym-summary-between-eigfunc-bar-PT}
\bar{\psi}_{n+1} (z) =
\frac{z^{n_{0}+1}}{ c_{n+1}}
 \Lambda_{-}^{-1}{\bar{\phi}^{(2)*}_{n_{0}-n}} (z^*)  .
\end{equation}

\subsubsection{Symmetries between the modified eigenfunctions}
Following from the above symmetry relations, we have
\begin{eqnarray}
\label{first-sym-summary-2c-N-and-M-PT}
 N_{n+1} (z)  &=&
\frac{1}{ c_{n+1}}
\Lambda_{-}{M^{*}_{n_{0}-n}} (z^*) ,\\
\label{first-sym-summary-2c-N-and-M-bar-PT}
\bar{N}_{n+1} (z) &=&
\frac{1}{ c_{n+1}}
\Lambda_{-}^{-1} {\bar{M}^{*}_{n_{0}-n}} (z^*).
\end{eqnarray}
We turn our attention next to compute the symmetries between the modified eigenfunctions defined by \eqref{mod_efns}. The symmetries given in (\ref{first-sym-summary-2c-N-and-M-PT}) and (\ref{first-sym-summary-2c-N-and-M-bar-PT}) induce symmetries between the ``primed" eigenfunctions given by
\begin{eqnarray}
\label{first-sym-summary-2c-N-and-M-prime-new-cond-comp-1b0}
 {N'_{n+1}} (z)   &=&  \tilde\Gamma\Lambda_{-}{M'^{*}_{n_{0}-n}} (z^*) ,\\
\label{first-sym-summary-2c-N-and-M-prime-new-cond-comp-2b2}
 {\bar{N}'_{n+1}} (z)      
&=&
  \tilde\Gamma\Lambda_{-}^{-1}{\bar{M}'^{*}_{n_{0}-n}} (z^*),
\end{eqnarray}
where we have used the identity $c^{*}_{n_{0}-n}c_{n+1}=c_{-\infty}$.

\subsubsection{Symmetries of the scattering data}
We start from (\ref{W-a}) combined with the symmetries between the eigenfunctions given in 
(\ref{first-sym-summary-between-eigfunc-PT}). We have
\begin{equation}
\label{W-a-3}
a(z) = \frac{ 1}{z^{n_{0}+1}} [\phi^{(1)}_{n+1} (z)  {\phi^{(1)*}_{n_{0}-n}} (z^*)  
            + \sigma \phi^{(2)}_{n+1} (z)    {\phi^{(2)*}_{n_{0}-n}} (z^*)  ]
          \;.
\end{equation}
Next, take the complex conjugate and let $z\rightarrow z^*,\;\; n\rightarrow n_{0}-n$ in (\ref{W-a-3}) to find the result
\begin{equation}
\label{discrete-scat-a-PT}
a(z) ~ = ~ a^*(z^*) .
\end{equation}
The derivation of the symmetry relation for the scattering data $\bar{a}(z)$ follows similar steps as for $a(z).$ Indeed, if one starts from the definition (\ref{W-a-bar}); utilize the symmetry between the eigenfunctions given in (\ref{first-sym-summary-between-eigfunc-bar-PT}), one arrives at the result
\begin{equation}
\label{discrete-scat-a-bar-3PT}
\bar{a}(z) = \bar{a}^*(z^*) .
\end{equation}
Note that the above two equations imply the following conclusion:
\begin{align*}
\label{eig-PT-conclusion}
&
 \text{If}~ a(z_{j})=0~\text{and}~ \bar{a}(\bar{z}_{j})=0,~\text{then}~ a(z_{j}^{*})=0~\text{and}~\bar{a}(\bar{z}_{j}^{*})=0 .
\end{align*}
Finally, we next determine the symmetry relationhip between $b(z)$ and $\bar{b}(z).$ As before, we start from 
Eq. (\ref{W-b-bar}) and use the symmetry between the eigenfunctions (\ref{first-sym-summary-between-eigfunc-PT}) to find
\begin{equation}
\label{W-b-bar-1PT}
\bar{b}(z) =
 \frac{1}{z^{n_{0}+1}} [\bar{\phi}^{(1)}_{n+1} (z) {\phi^{(1)*}_{n_{0}-n}} (z^*) 
            +\sigma  \bar{\phi}^{(2)}_{n+1} (z)  \phi^{(2)*}_{n_{0}-n} (z^*)   ] .
\end{equation}
With this at hand, it follows from (\ref{W-b}) that
\begin{equation}
\label{W-b-1-PT}
b(z)   =
  z^{n_{0}+1} [ \bar{\phi}^{(2)*}_{n_{0}-n} (z^*)     \phi^{(2)}_{n+1} (z) 
            +\sigma  \bar{\phi}^{(1)*}_{n_{0}-n} (z^*)  \phi^{(1)}_{n+1} (z)   ].
\end{equation}
Let $z\rightarrow z^*$ and $n_{0}-n\rightarrow n+1$ in (\ref{W-b-1-PT}), and complex conjugate the result to obtain the symmetry 
\begin{equation}
\label{discrete-scat-b-b-bar}
\bar{b}(z) = \frac{\sigma}{z^{2(n_{0}+1)}} b^*(z^*).
\end{equation}

\section{Inverse scattering and time dependence -- Discrete case}

\subsection{Inverse scattering: Riemann-Hilbert approach}
The next step of the inverse scattering transform is to solve an inverse problem with the goal of constructing an explicit formula for the potentials $Q_n(t)$ and $R_n(t)$ using the known scattering data. This is accomplished
by reformulating the AL scattering problem as a Riemann-Hilbert problem.
The scattering problem (\ref{disc-scat3}) can possess discrete eigenvalues that occur at the zeros of $a(z)$ and $\bar{a}(z)$, which will be denoted $\{ z_j :|z_j|>1\}_{j=1}^{J}$ and $\{ \bar{z}_j : |\bar{z}_j|<1\}_{j=1}^{\bar{J}}$, respectively (we assume that these zeros are simple and outside and inside the unit circle respectively). Furthermore, it is useful to define norming constants,
\begin{equation}
\label{norming}
    C_{j}\equiv\frac{b_{j}}{a_{z}(z_{j})},\;\;\bar C_{j}\equiv\frac{\bar b_j}{\bar a_{z}(\bar z_{j})},
\end{equation}
where $b_{j}$ and $\bar b_{j}$ are complex coefficients. The details of the Riemann-Hilbert problem for the AL coupled system are thoroughly discussed in \cite{APT_Book}. Presently, we quote the main result of the inverse problem. That is, from \eqref{rh1d} and \eqref{rh2d} it can be derived that the modified eigenfunctions satisfy a system of algebraic-integral equations involving the discrete eigenvalues, norming constants, and reflection coefficients:
 \begin{eqnarray}
\label{phi-linear-dependence-4-disc-prime-d-sys-1}
\bar{N}'_n(\bar{z}_j) 
&=&
\mathbf{\hat{e}_{1}}
+ 
 \sum_{\ell=1}^{J} C_{\ell} z_{\ell}^{-2n} \bigg[\frac{N_n' (z_{\ell})}{\bar{z}_j - z_{\ell}}
+ \frac{N_n' (-z_{\ell})}{\bar{z}_j + z_{\ell}} \bigg]
-
\frac{1}{2\pi i}\oint_{|\zeta| = 1}\frac{\zeta^{-2n}  \rho(\zeta)  N'_n(\zeta)}{\zeta -\bar{z}_j }d\zeta
,\\
\label{phi-linear-dependence-4-disc-prime-d-1-sys-3}
N'_n(z_j) 
&=&
\mathbf{\hat{e}_{2}}
+ 
 \sum_{\ell=1}^{\bar{J}} \bar{C}_{\ell} \bar{z}_{\ell}^{2n} 
 \bigg[\frac{\bar{N}_n' (\bar{z}_{\ell})}{z_j - \bar{z}_{\ell}}
+ \frac{\bar{N}_n' (-\bar{z}_{\ell})}{z_j + \bar{z}_{\ell}} \bigg]
+
\frac{1}{2\pi i}\oint_{|\zeta| = 1}\frac{\zeta^{2n}  \bar{\rho}(\zeta)  \bar{N}'_n(\zeta)}{\zeta- z_j }d\zeta
.
\end{eqnarray}
Once the eigenfunctions are determined from the above system, the potentials can be recovered using the following formulae derived using the asymptotic expansions of the modified eigenfunctions (see \cite{APT_Book} for details):
\begin{eqnarray}
\label{N-bar-prime-comp-2-Potential-Rn}
R_n
&=&
2  \sum_{j=1}^{J}  C_j z_j^{-2(n+1)}  N^{(2)'}_n (z_j)
+
\frac{1}{2\pi i}\oint_{|\zeta| = 1} \zeta^{-2(n+1)}  \rho(\zeta)  N^{(2)'}_n(\zeta)d\zeta
, \\
\label{N-bar-prime-comp-2-Potential-Qn}
Q_{n-1} 
&=&
- 2  \sum_{j=1}^{\bar{J}}  \bar{C}_j \bar{z}_j^{2(n-1)}  \bar{N}^{(1)'}_n (\bar{z}_j)
+
\frac{1}{2\pi i}\oint_{|\zeta| = 1} \zeta^{2(n-1)}  \bar{\rho}(\zeta)  \bar{N}^{(1)'}_n(\zeta)d\zeta
. 
\end{eqnarray}

\subsection{Time evolution}
In this section, we provide the time evolution of all scattering data and norming constants. Following similar lines of derivation as detailed in \cite{APT_Book} 
one finds the time-dependence of $a$ and $b$ is given by
\begin{equation}
\label{time-evolv-a-and-b-explicit}
a_{z}(z, t ) = 0\;,\;\;\;
b_{z}(z, t) = 2i\omega b(z,t).
\end{equation}
Similarly, the other set of scattering data satisfy 
\begin{equation}
\label{time-evolv-abar-and-bbar-explicit}
\bar{a}_{z}(z, t ) = 0\;,\;\;\;
\bar{b}_{z}(z, t ) =  -2i\omega  \bar{b}(z,t).
\end{equation}
For time-local cases, the explicit time evolution of the scattering data is thus given by
\begin{eqnarray}
&a(z,t)=a(z,0),\;\;&\bar{a}(z,t)=a(z,0),\\
&b(z,t)=e^{2i\omega t}b(z,0),\;\;&\bar{b}(z,t)=e^{-2i\omega t}\bar{b}(z,0).
\end{eqnarray}
For the time-shifted cases for which the initial scattering data is obtained at time $t=t_{0}/2$, the time evolution is given by
\begin{eqnarray}
&a(z,t)=a(z,t_{0}/2),\;\;&\bar{a}(z,t)=a(z,t_{0}/2)\\
&b(z,t)=e^{2i\omega (t-t_{0}/2)}b(z,t_{0}/2),\;\;&\bar{b}(z,t)=e^{-2i\omega (t-t_{0}/2)}\bar{b}(z,t_{0}/2),
\end{eqnarray}
The evolution of the norming constants $C_j$ and $\bar{C}_j$ defined in equation \eqref{norming}
is given for time-local cases by
\begin{equation}
\label{C-ell-evolve}
C_j(t)  
 = 
C_j (0) e^{2i\omega_{j} t},\;\;\; 
\bar{C}_j (t)
 =     \bar{C}_j (0) e^{-2i\bar{\omega}_{j} t},
\end{equation}
and for time-shifted cases by
\begin{equation}
\label{C-ell-evolve-shift}
C_j(t)  
 = 
C_j (t_{0}/2) e^{2i\omega_{j} (t-t_{0}/2)},\;\;\; 
\bar{C}_j (t)
 =     \bar{C}_j (t_{0}/2) e^{-2i\bar{\omega}_{j} (t-t_{0}/2)},
\end{equation}
where
\begin{equation}
 \label{omega-eig}
 \omega_j = \frac{1}{2}( z_j - z_j^{-1} )^2 , \;\;\;
 \bar{\omega}_j = \frac{1}{2}( \bar{z}_j - \bar{z}_j^{-1})^2 \;.
 \end{equation}

\section{Trace formulae and symmetries for $b_j, \bar{b}_j$ -- Discrete case}
For the time-shift reduction case (as well as the classical Ablowitz-Ladik case) the symmetries connect the scattering data and norming 
constants inside the unit circle to their corresponding quantities defined outside the unit circle. This implies that the eigenvalues
$z_j$ and norming constants $C_j$ are counted as free parameters and the values of $\bar{z}_j$ and $\bar{C}_j$ are uniquely determined by 
the underlying symmetries. However, in the cases involving spatial nonlocality, the symmetries of the scattering data and norming constants do {\it not} relate their respective values inside and outside of the unit circle. In order to understand the underlying symmetries of the norming constants 
$C_j=b_j/a_{z}(z_j)$ and $\bar{C}_j= \bar{b}_j/ \bar{a}_{z}(\bar{z}_j)$ we need to consider their numerators and denominators separately. For the denominators we employ trace formulae which can be used to show that the data  $a_{z}(z_j), \bar{a}_{z}(\bar{z}_j)$ depend on the eigenvalues $z_j, \bar{z}_j.$ For the numerators, it turns out that we can find symmetries relating $b_j, \bar{b}_j$. 

\subsection{Trace formulae}

\subsubsection{Space-time shifted reduction}
The derivation of the trace formulae follows an identical set of steps as shown in \cite{AM_discrete}. As such, the details are omitted. We 
assume that $a(z)$ and $\bar{a}(z)$ have simple zeros $\{\pm z_{j}: |z_{j}|>1\}_{j=1}^{J}$ and $\{\pm \bar{z}_{j}: |\bar{z}_{j}|<1\}_{j=1}^{J}$ respectively. 
The following trace formulae can be obtained:
\begin{eqnarray}
\label{Trace-a}
\log a(z)&\underset{|z|>1}{=}&\sum_{j=1}^{J} \log\bigg(\frac{z^{2}-z_{j}^{2}}{z^{2}-\bar{z}_{j}^{2}}\bigg) - \frac{1}{2\pi i}\oint_{|\zeta|=1}
\frac{\zeta\log [c_{-\infty} +  b(\zeta ) \bar{b}(\zeta ) ]}{\zeta^{2}-z^{2}}d\zeta,\\
\label{Trace-a-bar}
\log \bar{a}(z)&\underset{|z|<1}{=}&\sum_{j=1}^{J} \log\bigg(\frac{z^{2}-\bar{z}_{j}^{2}}{z^{2}-z_{j}^{2}}\bigg)+ \frac{1}{2\pi i}\oint_{|\zeta|=1}
\frac{\zeta\log [c_{-\infty} +  b(\zeta ) \bar{b}(\zeta ) ]}{\zeta^{2}-z^{2}}d\zeta.
\end{eqnarray}
We shall restrict the discussion to reflectionless potentials for which $b= \bar{b}=0.$ In this case, the second term in \eqref{Trace-a} vanishes, and the second term in \eqref{Trace-a-bar} becomes $\log c_{-\infty}$. If we assume we have only a single eigenvalue pair $z=\pm z_1, \pm\bar{z}_1 \in \mathbb{C}$, we find
\begin{equation}
\label{Trace-a-der-1}
a_{z}(z_1)= \frac{2z_1}{z_1^2 - \bar{z}_1^2}
 , \;\;\;\;\; 
 \bar{a}_{z}(\bar{z}_1)=   \frac{2 c_{-\infty}\bar{z}_1}{\bar{z}_1^2 - z_1^2} .
\end{equation}

\subsubsection{Space shifted reduction}
Again, the derivation follows the same set of steps as in the $\mathcal{PT}$-symmetric case in \cite{AM_discrete}, so the details are omitted. The difference from the space-time shifted case is the existence of more zeros of the scattering data $a(z)$ and $\bar{a}(z)$, i.e., they appear in quartets. 
Denote by
$\{\pm z_{j}, \pm z_{j}^{*}: |z_{j}|>1\}_{j=1}^{J}$ and $\{\pm\bar{z}_{j}, \pm \bar{z}_{j}^{*}: |\bar{z}_{j}|<1\}_{j=1}^{J}$, the (simple) zeros of
$a(z)$ and $\bar{a}(z)$ respectively. The general trace formulae are:
\begin{eqnarray}
\log a(z)&\underset{|z|>1}{=}&\sum_{j=1}^{J} \log \bigg[\frac{(z^{2}-z_{j}^{2})(z^{2}-z_{j}^{*2})}{(z^{2}-\bar{z}_{j}^{2})(z^{2}-\bar{z}_{j}^{*2})}\bigg]-
\frac{1}{2\pi i}\oint_{|\zeta|=1}\frac{\zeta\log[c_{-\infty} +  b(\zeta)\bar{b}(\zeta)]} {\zeta^{2}-z^{2}}d\zeta,\\
\log\bar{a}(z)&\underset{|z|<1}{=}&\sum_{j=1}^{J} \log\bigg[\frac{(z^{2}-\bar{z}_{j}^{2})(z^{2}-\bar{z}_{j}^{*2})}{(z^{2}-z_{j}^{2})(z^{2}-z_{j}^{*2})}\bigg]+\frac{1}{2\pi i}\oint_{|\zeta|=1}\frac{\zeta\log[c_{-\infty} +b(\zeta)\bar{b}(\zeta)] }{\zeta^{2}-z^{2}}d\zeta.
\end{eqnarray}
Again, we consider reflectionless potentials and assume a single complex eigenvalue quartet \\
$z=\pm z_1, \pm z_1^{*},\pm\bar{z}_1, \pm\bar{z}_1^{*}$, in which case we find
\begin{eqnarray}
\label{Trace-a-der-PT-eig1}
a_{z}(z_1) =  \frac{2z_1(z_1^2-z_1^{*2})}{(z_1^2-\bar{z}_1^{2})(z_1^2-\bar{z}_1^{*2})}
,\;\;\;a_{z}(z^*_1) =  \frac{2z^*_1(z_1^{*2}-z_1^2)}{(z_1^{*2} - \bar{z}_1^{2})(z_1^{*2}-\bar{z}_1^{*2})}
,\\
\label{Trace-a-ba-der-PT-eig1}
\bar{a}_{z}(\bar{z}_1) =  
\frac{2 c_{-\infty}\bar{z}_1(\bar{z}_1^2-\bar{z}_1^{*2})}{(\bar{z}_1^2 - z_1^2)(\bar{z}_1^2 - z_1^{*2})}
,\;\;\;\bar{a}_{z}(\bar{z}^*_1) =  
\frac{2 c_{-\infty}\bar{z}^*_1(\bar{z}_1^{*2}-\bar{z}_1^{2})}{(\bar{z}_1^{*2} - z_1^2)(\bar{z}_1^{*2} - z_1^{*2})}
.
\end{eqnarray}
Later, for simplicity, we will consider the case when $z_{1},\bar{z}_{1}$ are {\it real} (this is equivalent to restricting the discrete eigenvalues to the imaginary axis in the continuous case). With this restriction, the calculations become identical to the space-time shift case, and the formulae for $a_{z}$ and $\bar{a}_{z}$ given in (\ref{Trace-a-der-1}) hold.

\subsection{Computing symmetries of $b_j$ and $\bar{b}_j$}
In this section we obtain the symmetries that $b_j$ and $\bar{b}_j$ satisfy. 
This in turn will be later used to determine the dependence of the norming constants $C_j$ and $\bar{C}_j$ on these eigenvalues. 
In the calculations that follow we will make frequent use of the fact established in \cite{APT_Book} that  
$c_{-\infty}= z_1^2/\bar{z}^2_1$ in the $J=1$ reflectionless case. 

\subsubsection{Space-time shifted reduction}
At a discrete eigenvalue $z_{j}$, we have the relationships
\begin{equation}
\label{phi-linear-dependence-1-disc-eig-Jost-eig-2}
M'_{n} (z_j,t) = b_j(t) z_j^{-2n} N'_{n} (z_j,t) .
\end{equation}
Using the symmetry relation \eqref{sym-prime-RST-1}, together with the second component of Eq.~(\ref{phi-linear-dependence-1-disc-eig-Jost-eig-2}), results in
\begin{equation}
\label{first-sym-summary-2c-N-and-M-prime-new-cond-comp-1b1-b}
 {N^{(1)'}_{n+1}} (z_j,t)  = -\sigma  c_{-\infty}^{-1}  b_j(t_{0}-t) z_j^{-2(n_{0}-n)} N^{(2)'}_{n_{0}-n} (z_j,t_{0}-t) .
 \end{equation}
 Next, applying the symmetry relation
 (\ref{first-sym-summary-2c-N-and-M-new}) to the first component of
(\ref{phi-linear-dependence-1-disc-eig-Jost-eig-2}) gives rise to
 \begin{equation}
\label{first-sym-summary-2c-N-and-M-prime-new-cond-comp-2b-b}
  {N^{(2)'}_{n_{0}-n}} (z_j,t_{0}-t)     =  b_j(t) z_j^{-2(n+1)} N^{(1)'}_{n+1} (z_j,t) .
\end{equation}
Substituting Eq.\eqref{first-sym-summary-2c-N-and-M-prime-new-cond-comp-2b-b} into \eqref{first-sym-summary-2c-N-and-M-prime-new-cond-comp-1b1-b} one 
obtains the symmetry 
\begin{equation}
\label{b-sym-at-eig-val-RST}
 b_j(t_{0}-t) b_j(t)
 =
 -\sigma c_{-\infty} z_j^{2(n_{0}+1)} \;
\;\;\;\;\; \Longrightarrow \;\;\;\;\;
 b^2_j(t_{0}/2) = -\sigma c_{-\infty} z_j^{2(n_{0}+1)}
  ,
 \end{equation} 
where we used the time evolution of $b(t)$ given by Eq.~(\ref{time-evolv-a-and-b-explicit}). For a one soliton solution we find
 \begin{equation} 
 \label{b1-one-sol}
b_1(0) =  \gamma_{1}e^{i (1+\sigma )\pi /4}\frac{z_1^{n_{0}+2}}{\bar{z}_1} \;, \;\;\; \gamma_{1} = \pm 1.
 \end{equation} 
To determine the value of $\bar{b}_j(t)$, one follows similar steps, which lead to
\begin{equation}
\label{b-b-bar-RST-2} 
 \bar{b}_j (t_{0}-t) \bar{b}_j (t) = - \sigma  c_{-\infty}  \bar{z}_j^{-2(n_{0}+1)} \;\;\;\; \Longrightarrow \;\;\;\;  
\bar{b}^2_j (t_{0}/2) = - \sigma  c_{-\infty} \bar{z}_j^{-2(n_{0}+1)} .
\end{equation}
In obtaining the expression for $\bar{b}_j (t_{0}/2)$ we made use of the evolution of the scattering data $\bar{b}_j (t)$ given 
by Eq.~(\ref{time-evolv-abar-and-bbar-explicit}). For the one soliton case we find
 \begin{equation} 
 \label{b1-bar-one-sol}
\bar{b}_1(t_{0}/2) =   \bar\gamma_{1}e^{i (1+\sigma )\pi /4}\frac{z_1}{\bar{z}_1^{n_{0}+2}} \;,\;\;\; \bar\gamma_{1}= \pm 1 .
 \end{equation} 
With the help of Eq.~(\ref{Trace-a-der-1}) we are now ready to compute the norming constants $C_1$ and $\bar{C}_1$. 
Including the time evolution of the norming constants we have 
\begin{eqnarray}
\label{norming-const-C1-C1-bar-final}
C_1(t) &=&  \gamma_{1}e^{i (1+\sigma )\pi /4}(z_1^2 - \bar{z}_1^2)\frac{z_1^{n_{0}+1} }{2 \bar{z}_1}e^{-i\omega_{1}t_{0}}e^{2i\omega_{1}t} 
 ,\\
 \label{norming-const-C1-C1-bar-final_2}
\bar{C}_1(t) &=&  \bar\gamma_{1}e^{i (1+\sigma )\pi /4}(\bar{z}_1^2 - z_1^2)\frac{1}{2 z_1 \bar{z}_1^{n_{0}+1}}e^{i\bar\omega_{1}t_{0}}e^{-2i\bar\omega_{1}t} .
\end{eqnarray}

\subsubsection{Space shifted reduction}
Our starting point is 
again Eq.~(\ref{phi-linear-dependence-1-disc-eig-Jost-eig-2}). Substituting the symmetry condition 
(\ref{first-sym-summary-2c-N-and-M-prime-new-cond-comp-1b0}) into the second component one finds
\begin{equation}
\label{first-sym-summary-2c-N-and-M-prime-new-cond-comp-1b1-b-PT}
 {N^{(1)'}_{n+1}} (z_j)  = -\sigma c_{-\infty}^{-1} ~  b_j^{*} z_j^{-2(n_{0}-n)} N^{(2){'*}}_{n_{0}-n} (z_j^{*}) .
 \end{equation}
Next, we use the symmetry (\ref{first-sym-summary-2c-N-and-M-prime-new-cond-comp-1b0})
to rewrite the first component of Eq.~(\ref{phi-linear-dependence-1-disc-eig-Jost-eig-2}) in the form
 \begin{equation}
\label{first-sym-summary-2c-N-and-M-prime-new-cond-comp-2b-b-PT}
  {N^{(2)'*}_{n_{0}-n}} (z_j^{*})     =  b^*_j z_j^{-2(n+1)} N^{(1){'}}_{n+1} (z_j)  .
\end{equation}
Substituting Eq.~(\ref{first-sym-summary-2c-N-and-M-prime-new-cond-comp-2b-b-PT}) back 
into (\ref{first-sym-summary-2c-N-and-M-prime-new-cond-comp-1b1-b-PT}) gives
\begin{equation}
\label{b1-t-PT}
|b_j|^2   = -\sigma c_{-\infty}z_j^{2(n_{0}+1)}    .
 \end{equation}
For a one soliton solution 
(recall $c_{-\infty}= z_1^2/\bar{z}^2_1$)
\begin{equation}
|b_1|^2   = -\frac{\sigma z_1^{2(n_{0}+2)}}{\bar{z}_1^2}.
 \end{equation}
This is only possible if $\sigma=-1$ and $z_{1},\bar{z}_{1}\in\mathbb{R}$, in which case we have
\begin{equation}
    \label{b1-t-PT-1}
    b_{1}=e^{i\theta_{1}}\frac{z_{1}^{n_{0}+2}}{\bar{z}_{1}},\;\;\;\theta_{1}\in\mathbb{R}.
\end{equation}
A similar expression can be obtained for $\bar{b}_j.$ Indeed, following an analogous procedure one can find
\begin{equation}
\label{b1-t-PT-bar}
|\bar{b}_j |^2   = -\sigma  c_{-\infty}\bar{z}_j^{-2(n_{0}+1)} 
   .
 \end{equation}
 Again, for a one soliton solution we take $\sigma=-1$ and $z_{1},\bar{z}_{1}\in\mathbb{R}$, leading to
\begin{equation}
\label{b1-t-PT-bar-2}
\bar{b}_1   = e^{i\bar{\theta}_1}\frac{z_1}{\bar{z}^{n_{0}+2}_1}   \;,\;\;\; \bar{\theta}_{1}\in\mathbb{R}.
 \end{equation}
To write down a closed form expression for the norming constants $C_1$ and $\bar{C}_1$, we use the above formulae, the definition and time-evolution of the norming 
constants, and Eq.~(\ref{Trace-a-der-1}):
\begin{eqnarray}
\label{norming-const-C1-C1-bar-final-PT}
C_1(t) &=&  e^{i \theta_1}(z_1^2 - \bar{z}_1^2)\frac{z_1^{n_{0}+1} }{2 \bar{z}_1} e^{2i\omega_{1}t}
 ,\\
 \label{norming-const-C1-C1-bar-final-PT-2}
\bar{C}_1(t) &=&  e^{i \bar{\theta}_1}(\bar{z}_1^2 - z_1^2)\frac{1}{2 z_1 \bar{z}_1^{n_{0}+1}}e^{-2i\bar\omega_{1}t} .
\end{eqnarray}

\section{1-soliton solutions -- Discrete case}
\label{one-soliton} 
In this section we compute 1-soliton solutions for the time, space-time, and space shifted discrete NLS systems. Before considering the relevant nonlocal reductions, one can obtain the general 1-soliton for the coupled $(Q_n,R_n)$ system by setting $J=\bar{J}=1$ and $\rho=\bar\rho=0$ in \eqref{phi-linear-dependence-4-disc-prime-d-sys-1}-\eqref{phi-linear-dependence-4-disc-prime-d-1-sys-3}, which are simply an algebraic system of equations under these assumptions. Once the relevant 
eigenfunctions $\bar{N}^{(1)'}_n(\bar{z}_1,t)$ and $N^{(2)'}_n(z_1,t)$ are known, they may be substituted into \eqref{N-bar-prime-comp-2-Potential-Rn} and \eqref{N-bar-prime-comp-2-Potential-Qn} to find:
\begin{eqnarray}
\label{N-bar-prime-comp-2-Potential-Rn-sol}
R_n (t )
&=&  
\frac{   2   C_1(t ) z_1^{-2(n+1)}   (z^2_1-\bar{z}^2_1 )^2}
{(z^2_1-\bar{z}^2_1)^2 + 4 \bar{C}_1(t ) C_1(t ) \bar{z}_1^{2n+2}  z_1^{-2n}} 
,\\
\label{N-bar-prime-comp-2-Potential-Qn-11}
Q_{n} (t )
&=&
 \frac{ - 2  \bar{C}_1(t ) \bar{z}_1^{2n}   (z^2_1-\bar{z}^2_1 )^2}
{(z^2_1-\bar{z}^2_1 )^2+4 \bar{C}_1(t ) C_1(t ) \bar{z}_1^{2n+2}  z_1^{-2n}}
 .
\end{eqnarray}
To determine the 1-soliton solutions for each specific reduction, we apply the relevant symmetries in scattering space to the general form \eqref{N-bar-prime-comp-2-Potential-Qn-11}.
\subsection{Time shifted reduction}
In this case, we have symmetries connecting the ``inside" scattering data to the ``outside" scattering data. Particularly, from \eqref{eig-sym} we have $\bar{z}_1=1/z_1$ with $z_1\in\mathbb{C}$ and from \eqref{discrete-scat-a-3} and \eqref{discrete-scat-b-b-bar-RT} together with the definition of the norming constants \eqref{norming} we obtain
\begin{equation}
\label{time-shift-norming-sym}
\bar{C}_1 (t)  = -\sigma z_1^{-2} C_1(t_{0}-t )=  -\sigma z_1^{-2}   C_1 (t_{0}/2) e^{i\omega_{1}t_{0}}e^{-2i\omega_1 t} .
 \end{equation} 
Substituting \eqref{time-shift-norming-sym} into
(\ref{N-bar-prime-comp-2-Potential-Qn-11}) gives
\begin{equation}
\label{Qn-RT-NLS-sol}
Q_{n} (t )
=
\frac{2\sigma C_{1}(t_{0}/2)e^{i\omega_{1}t_{0}}e^{-2i\omega_{1}t}{z}_{1}^{-2(n+1)}}{1-\sigma\frac{4C_{1}(t_{0}/2)^{2}}{(z_{1}^{2}-{z}_{1}^{-2})^{2}}z_{1}^{-4(n+1)}} .
\end{equation}
This soliton solution has two arbitrary parameters $z_{1}$ and $C_{1}(t_{0}/2)$ (recall $\omega_{1}=\frac{1}{2}(z_{1}-z_{1}^{-1})^{2}$) which in general may be complex. We note that for certain initial conditions, the denominator of \eqref{Qn-RT-NLS-sol} can vanish (for all time); and as such, these should be excluded from the set of admissible initial conditions.
\subsection{Space-time shifted reduction}
Recall that unlike the time-shifted case, here the ``inside" and ``outside" scattering data are not connected. Instead, $z_{1}$ and $\bar{z}_{1}$ are free parameters and the norming constants can be expressed explicitly in terms of these two parameters using the formulae given in \eqref{norming-const-C1-C1-bar-final}-\eqref{norming-const-C1-C1-bar-final_2}. Substituting these into (\ref{N-bar-prime-comp-2-Potential-Qn-11}), we obtain the 1-soliton solution
\begin{equation}
\label{N-bar-prime-comp-2-Potential-Qn-11-new-RST}
Q_{n}(t)=\frac{e^{i(1+\sigma)\pi/4}(z_{1}\bar{z}_{1})^{-1}(z_{1}^{2}-\bar{z}_{1}^{2})\bar{z}_{1}^{2n}e^{-2i\bar\omega_{1}t}}{\bar\gamma_{1}\bar{z}_{1}^{n_{0}}e^{-i\bar\omega_{1}t_{0}}+\sigma\gamma_{1}z_{1}^{n_{0}}e^{-i\omega_{1}t_{0}}z_{1}^{-2n}\bar{z}_{1}^{2n}e^{2i(\omega_{1}-\bar\omega_{1})t}}
,
\end{equation}
where we have used the fact that $e^{i(1+\sigma)\pi/2}=-\sigma$. The parameters $z_1, \bar{z}_1$ are free complex constants defined outside/inside the unit circle respectively, and $\gamma_{1}, \bar{\gamma}_{1}$ can be chosen arbitrarily as 
either $\pm1$.
We remark that while no choice of parameters will cause the denominator to vanish for all time, this soliton can develop singularity in finite time.
\subsection{Space shifted reduction}
While the previous two reductions allow for 1-soliton solutions for either choice of $\sigma$ and complex eigenvalues, in the space shifted case we have the restrictions that $\sigma=-1$ and $z_{1},\bar{z}_{1}\in\mathbb{R}$ (see Eq.\eqref{b1-t-PT-1}). With this, substituting \eqref{norming-const-C1-C1-bar-final-PT}-\eqref{norming-const-C1-C1-bar-final-PT-2} into \eqref{N-bar-prime-comp-2-Potential-Qn-11} gives
\begin{equation}
\label{Qn-RST-NLS-sol-left-final2}
Q_{n}(t)=\frac{(z_{1}\bar{z}_{1})^{-1}(z_{1}^{2}-\bar{z}_{1}^{2})\bar{z}_{1}^{2n}e^{-2i\bar\omega_{1}t}}{e^{-i\bar\theta_{1}}\bar{z}_{1}^{n_{0}}-e^{i\theta_{1}}z_{1}^{n_{0}}z_{1}^{-2n}\bar{z}_{1}^{2n}e^{2i(\omega_{1}-\bar\omega_{1})t}}.
\end{equation}
This soliton has four arbitrary real parameters $z_{1},\bar{z}_{1},\theta_{1},\bar\theta_{1}$, and like the space-time shifted case, can develop singularity in finite time.

\section{Conclusion}

In this work we have presented the inverse scattering transform for both continuous and discrete examples of nonlocal integrable systems possessing a shifted spatial and/or temporal argument. The spatial shifting parameter appears as an additional factor in the scattering data, while the primary role of the temporal shifting parameter is a restriction on the time at which the initial condition must be provided. We have shown the details of the IST for each case in an effort to illustrate the effect of the shifting parameters throughout. Additionally, this work serves the purpose of compiling the symmetries and soliton solutions for a wide variety of nonlocal integrable equations into a single reference.

Many directions for future exploration remain in the area of nonlocal integrable systems. The Davey-Stewartson equation \cite{DS} also has integrable space and/or time shifted variants, for example
    \begin{equation}
        \begin{cases}
            iq_{t}(x,y,t)+\frac{1}{2}[\gamma^{2}q_{xx}(x,y,t)+q_{yy}(x,y,t)]=[\varphi(x,y,t)-q(x,y,t)q(x_{0}-x,y_{0}-y,t_{0}-t)]q(x,y,t) \;, \\
            \varphi_{xx}(x,y,t)-\gamma^{2}\varphi_{yy}(x,y,t)=2[q(x,y,t)q(x_{0}-x,y_{0}-y,t_{0}-t)]_{xx},\;\;\;\gamma^{2}=\pm1 \;.
        \end{cases}
    \end{equation}
The IST for such nonlocal (2+1)D systems has not yet been studied, and will be explored in a future work. Another interesting direction would be to consider NLS-type equations with multiple different types of shifted nonlocality, which can be obtained from matrix-valued versions of the integrable coupled NLS system. The simplest example arises from the 2-component vector version of the system \eqref{q-sys}-\eqref{r-sys}:
\begin{eqnarray}
    iq^{(1)}_{t}(x,t)&=&q^{(1)}_{xx}(x,t)-2[q^{(1)}(x,t)r^{(1)}(x,t)+q^{(2)}(x,t)r^{(2)}(x,t)]q^{(1)}(x,t)\\
    iq^{(2)}_{t}(x,t)&=&q^{(2)}_{xx}(x,t)-2[q^{(1)}(x,t)r^{(1)}(x,t)+q^{(2)}(x,t)r^{(2)}(x,t)]q^{(2)}(x,t)\nonumber\\
    -ir^{(1)}_{t}(x,t)&=&r^{(1)}_{xx}(x,t)-2[q^{(1)}(x,t)r^{(1)}(x,t)+q^{(2)}(x,t)r^{(2)}(x,t)]r^{(1)}(x,t)\nonumber\\
    -ir^{(2)}_{t}(x,t)&=&r^{(2)}_{xx}(x,t)-2[q^{(1)}(x,t)r^{(1)}(x,t)+q^{(2)}(x,t)r^{(2)}(x,t)]r^{(2)}(x,t)\nonumber
\end{eqnarray}
Following the technique of \cite{Ma} used to obtain similar unshifted nonlocal equations, making the set of reductions $r^{(1)}(x,t)=q^{(1)}(x,t_{0}-t)$, $r^{(2)}(x,t)=q^{(2)}(x,t_{0}-t)$, and $q^{(2)}(x,t)=q^{(1)}(x_{0}-x,t)$, one obtains the following integrable evolution equation for the function $q^{(1)}(x,t) \equiv q(x,t)$
involving terms with space, time, and space-time shifts:
\begin{equation}
    iq_{t}(x,t)=q_{xx}(x,t)-2[q(x,t)q(x,t_{0}-t)+q(x_{0}-x,t)q(x_{0}-x,t_{0}-t)]q(x,t).
\end{equation}
Understanding the scattering-space symmetries and soliton solutions of equations such as this could be a topic for further study.\\

\section*{Acknowledgements}
This paper is dedicated to our long time esteemed colleague Dave Kaup. MJA was partially supported by  NSF under grants DMS-2005343 and DMS-2306290.

\end{document}